\documentclass[aps,prd,twocolumn,groupedaddress,nofootinbib]{revtex4}

\usepackage[scientific-notation=false]{siunitx}
\usepackage {natbib}
\usepackage {graphicx}
\usepackage {color}

\usepackage{amsmath, amsfonts, amssymb}
\usepackage{hyperref}

\usepackage{subfigure}

\usepackage{float}
\usepackage{newfloat,algpseudocode}
\usepackage[size=small]{caption}
\usepackage{etoolbox}
\usepackage[normalem]{ulem}

\AtEndEnvironment{algorithm}{\noindent\hrulefill\par\nobreak\vskip+8pt}

\DeclareFloatingEnvironment[
    fileext=loa,
    listname=List of Algorithms,
    name=ALG.,
    placement=tbhp,
]{algorithm}
\DeclareCaptionFormat{algorithms}{\vskip-15pt\hrulefill\par#1#2#3\vskip-6pt\hrulefill}
\algblock[Input]{Input}{EndInput}
\algblockdefx[Input]{Input}{EndInput}%
    [1]{\textbf{Input} #1}%
    {}
\algblock[Output]{Output}{EndOutput}
\algblockdefx[Output]{Output}{EndOutput}%
    [1]{\textbf{Output} #1}%
    {}

\newcommand{\mF}{\mathcal{F}} 
\newcommand{\mG}{\mathcal{G}} 
\newcommand{\mS}{\mathcal{S}} 

\newcommand{\dop}{\lambda}

\newcommand{\dopS}{\mathbb{P}}
\newcommand{\dopT}{\mathbb{T}}

\newcommand{\CC}{C}	

\newcommand{\CCtot}{\CC_{\mathrm{tot}}}

\newcommand{\hrms}{h_{\mathrm{rms}}}

\newcommand{\hnot}{h_{0}}

\newcommand{\intd}[1]{\mathrm{d}#1\,}

\newcommand{\FA}{\mathrm{fA}}
\newcommand{\FD}{\mathrm{fD}}

\newcommand{\pFA}{p_{\FA}}
\newcommand{\pFD}{p_{\FD}}

\newcommand{\thresh}{\mathrm{th}}

\newcommand{\Nseg}{N}

\newcommand{\Tseg}{\Delta T}	
\newcommand{\Tobs}{T}
\newcommand{\Tdata}{T_{\mathrm{data}}}
\newcommand{\Tspan}{T_{\mathrm{span}}}

\newcommand{\Nt}{\mathcal{N}} 	

\newcommand{\co}[1]{\widetilde{#1}}
\newcommand{\ic}[1]{\widehat{#1}}

\newcommand{\SFT}{\mathrm{SFT}}

\newcommand{\opt}[1]{{#1}_{\mathrm{opt}}}

\newcommand{\refac}{\gamma}
\newcommand{\vol}{\mathcal{V}}

\newcommand{\avg}[1]{\left\langle #1 \right\rangle}

\newcommand{\Sn}{S_{\mathrm{n}}}

\newcommand{\SNR}{\mathrm{SNR}}

\newcommand{\nc}{{\co{n}}}
\newcommand{\nf}{{\ic{n}}}

\newcommand{\abs}[1]{|#1|}

\newcommand{\mismax}{m}

\newcommand{\KWS}{{\mathrm{KWS}}}

\newcommand{\thick}{\theta}

\newcommand{\erfc}{\mathrm{erfc}}

\newcommand{\pacfactor}{\xi}

\newcommand{\days}{\mathrm{d}}
\newcommand{\Days}{\mathrm{days}}
\newcommand{\Hz}{\mathrm{Hz}}

\newcommand{\years}{\mathrm{y}}

\newcommand{\secs}{\mathrm{s}}

\newcommand{\Nsft}{N_{\mathrm{SFT}}}
\newcommand{\Tsft}{T_{\mathrm{SFT}}}
\newcommand{\Ndet}{N_\mathrm{det}}

\newcommand{\seglist}{\mathcal{L}}
\newcommand{\gudlist}{\mathcal{G}}
\newcommand{\sftlist}{\mathcal{D}}
\newcommand{\gudcost}{\mathcal{X}}

\newcommand{\sensdep}{\mathcal{D}}

\def\commitID{commitID: 159c81559e868171391182f0633cca08cefeccd2}
\def\commitDATE{ Sun Apr 3 19:11:57 2016 +0200}

\def\commitSTATUS{CLEAN}

\newcommand{\dcc}{LIGO-P1500178-v3}

\newcommand{\ltitle}{Optimizing the StackSlide setup and data selection for continuous-gravitational-wave searches in realistic detector data}
\newcommand{\stitle}{Optimizing the StackSlide setup and data selection for continuous-gravitational-wave searches in realistic detector data}

\hypersetup{
  pdftitle={\ltitle},
  pdfauthor={M. Shaltev},
  pdfkeywords={\$\commitID-\commitSTATUS{} \$}
}

\begin{document}

\title[\stitle]{\ltitle}
\author{M.\ Shaltev}
\affiliation{Albert-Einstein-Institut, Callinstr.\ 38, 30167 Hannover, Germany}
\date{\commitDATE\\\mbox{\dcc}}

\begin{abstract}
The search for continuous gravitational waves in a wide parameter space 
at fixed computing cost is most efficiently done with semicoherent methods, 
e.g. StackSlide, due to the prohibitive computing cost of the fully coherent 
search strategies. Prix\&Shaltev~\cite{PrixShaltev2011} have
developed a semi-analytic method for finding \emph{optimal} StackSlide parameters at
fixed computing cost under ideal data conditions, i.e. gap-less data and constant
noise floor. In this work we consider more realistic conditions by allowing for gaps in the
data and changes in noise level. We show how the sensitivity optimization can be
decoupled from the data selection problem. To find optimal 
semicoherent search parameters we apply a numerical optimization using as
example the semicoherent StackSlide search. 
We also describe three different data selection algorithms.
Thus the outcome of the numerical optimization consists of the optimal search
parameters and the selected dataset. We first test the numerical optimization procedure under ideal
conditions and show that we can reproduce the results of the analytical method.
Then we gradually relax the conditions on the data and find that a compact data 
selection algorithm yields higher sensitivity compared to a greedy data selection 
procedure.
\end{abstract}

\pacs{XXX}

\maketitle

\section{Introduction}

The enormous computational requirement of the wide parameter-space
searches for continuous gravitational waves impose a cautious use of the
available computing resources, as we always aim at maximal sensitivity
\footnote{Given unlimited computing power and / or a targeted search, i.e., when the sky position 
and frequency evolution of the source are known, we would prefer coherent matched-filtering
technique, see \cite{Jaranowski:1998qm}. In all other cases however, semicoherent searches may yield better
results in terms of sensitivity, see \cite{Brady:1998nj,PhysRevD.72.042004,PrixShaltev2011}.}. 
In this respect to maximize the sensitivity of a semicoherent search, 
for example StackSlide \cite{Brady:1998nj,PhysRevD.72.042004},
means that we need to choose the optimal search parameters, namely
number of segments, segment duration and  optimal maximal  
mismatch on the coarse and fine grid.  How to do this at fixed computing cost 
for a single astrophysical target and under the ideal conditions of constant noise 
floor and data without gaps has been studied in \cite{PrixShaltev2011}, where 
analytical expressions have been derived to determine the optimal search parameters.
In Ref. \cite{Ming:2015jla}, while still assuming ideal conditions on the data, a framework 
to distribute the total available computing cost between different possible targets 
based on astrophysical priors has been developed. However under realistic conditions 
the available data, as collected for 
example from the Laser Interferometer Gravitational-Wave Observatory (LIGO) 
\cite{Abbott:2007kv,Aasi:2014mqd}, can be fragmented, e.g., due to down-time of 
the detectors, and there might be variations in the noise floor. 
The fragmentation of the data can significantly affect the
 computing cost function and thus the optimal search parameters. On the
 other  hand, the noise fluctuations suggest the use of a data selection
 procedure in  order to spend the available  computing cycles searching
 data of higher quality.

In this work we extend Ref. \cite{PrixShaltev2011} to these more realistic conditions
 by taking into account possible gaps in the data and noise level changes.  First
 we show, how the real conditions manifest in the sensitivity function. Then 
we reformulate the problem, such that a numerical optimization procedure can be 
applied to find optimal semicoherent search parameters. We also describe a 
suitable data selection algorithm.
The outcome of the proposed numerical optimization are the optimal search
 parameters and the selected data, so that the search can be performed in practice.
 We first test the numerical optimization procedure under ideal
 conditions and obtain the results of the analytical method proposed in 
 Ref. \cite{PrixShaltev2011}. Then we give examples of practical application.

This paper is organized as follows. In Sec. \ref{sec:2} we introduce the
 ingredients of the search-optimization method, i.e., the threshold 
signal-to-noise ratio (SNR), the sensitivity  function,  and the computing
cost function. The numerical optimization of the search parameters and 
in particular the data selection method are described in Sec.  \ref{sec:3}. 
In Sec. \ref{sec:4} we give examples of practical application
and discuss in Sec. \ref{sec:5}.

\subsubsection*{Notation}
We use tilde when referring to fully coherent quantities, $\co{Q}$, 
and overhat when referring to semicoherent quantities, $\ic{Q}$. 

\section{Threshold SNR, sensitivity function and computing cost}
\label{sec:2}

In this section we introduce the main ingredients needed to 
define the optimization problem, i.e., to find the number of segments
$\Nseg$ with segment duration $\Tseg$ and coarse and fine grid
mismatch, $\co{m}$ resp. $\ic{m}$, which maximize the sensitivity of 
the search at fixed computing cost $\CC_{0}$. These ingredients are the
 threshold SNR required for detection, the sensitivity function, which we 
want to maximize, and the  computing cost function.

\subsection{Threshold SNR}

A claim for detection when searching for a signal, in particular a weak signal,
 in the presence of noise is sensible only in the context of two well defined 
quantities. The first one, called false-alarm probability, is the probability
to falsely claim a detection when the signal is not present in the data. The second
quantity, called false-dismissal probability, is the probability to miss the 
signal even if the signal is indeed present in the data.

When a signal is present in the data the semicoherent StackSlide statistic follows a 
non-central $\chi^{2}$ distribution with $4\Nseg$ degrees of freedom. 
Using the definitions  of  \cite{PrixShaltev2011}, we
 denote this by $\chi^{2}_{4\Nseg}(\hat{\mF},\hat{\rho}^{2})$, where 
$\hat{\rho}^2$ is the non-centrality parameter, i.e., the sum 
of the squared SNR of the individual segments
$\hat{\rho}^2\equiv\sum_{i=1}^{\Nseg}\co{\rho}^{2}_{i}$. We can
 integrate the false-alarm probability 
\begin{equation}
 \label{eq:1}
\pFA(\hat{\mF}_\thresh) = \int_{\hat{\mF}_\thresh}^{\infty}\intd{\hat{\mF}}\chi_{4\Nseg}^{2}(\hat{\mF};0)\ ,
\end{equation}
and by inversion for a given false-alarm $\pFA^{*}$ obtain the threshold
 $\hat{\mF}_\thresh$.
For a pre-defined false-dismissal $\pFD^{*}$ probability
\begin{equation}
 \label{eq:2}
\pFD(\hat{\mF}_\thresh,\hat{\rho}^{2}) = \int_{-\infty}^{\hat{\mF}_\thresh}\intd{\hat{\mF}} \chi^{2}_{4\Nseg}(\hat{\mF};\hat{\rho}^{2})\ ,
\end{equation}
using $\hat{\mF}_{\thresh}$  we aim to obtain the critical non-centrality
\begin{equation}
 \label{eq:3}
\hat{\rho}^{*2}=\hat{\rho}^{2}(\pFA^{*},\pFD^{*},\Nseg)\ , 
\end{equation}
and thus the threshold SNR.

The computation of  the critical non-centrality $\hat{\rho}^{*2}$  is
complicated
by the fact that for wide parameter-space searches the right-hand
side of Eq. \eqref{eq:2} requires averaging  over sky position and polarization parameters
of the signal carried out at fixed intrinsic amplitude $\hnot$.  In Ref. \cite{PrixShaltev2011} for example,
a signal population of constant SNR has been assumed. Therefore by application of 
the central limit theorem and approximation
of the $\chi^{2}$ distribution by a Gaussian distribution, Eq. \eqref{eq:2} has
been analytically integrated and inverted to obtain \eqref{eq:3}.
For weak signals and large number of segments (see Fig. 2 in Ref. \cite{PrixShaltev2011}) 
this results in the  
``weak-signal Gauss (WSG) approximation'' for the critical non-centrality parameter
\begin{equation}
 \label{eq:4}
\hat{\rho}^{*2}(\pFA^{*},\pFD^{*},\Nseg)\approx2\sqrt{4\Nseg}(\erfc^{-1}(2\pFA^{*})+\erfc^{-1}(2\pFD^{*}))\ .
\end{equation}
With this we define the per-segment threshold SNR
\begin{equation}
 \label{eq:5}
\rho^{*} \equiv\sqrt{\frac{\hat{\rho}^{*2}(\pFA^{*},\pFD^{*},\Nseg)}{\Nseg}}\ .
\end{equation}

Recently a new semi-analytical method has been
 developed to estimate the sensitivity of a search \cite{PhysRevD.85.042003}.
 In this method the assumption of signal distribution of constant SNR has been relaxed,
where a semi-analytical approximation  for the computation of an isotropic threshold SNR
 has been introduced. We refer to this method as the KWS approximation. 
In the KWS approximation the averaged over segments threshold $\rho^{*}$ is obtained
 recursively. At iteration $i$ the value of $\rho^{*}$ is
\begin{equation}
 \label{eq:6}
\rho=F\big(\rho\big)\ ,
\end{equation}
where 
\begin{equation}
\label{eq:7}
 \rho_{i+1}=F\left(\frac{\rho_{i}+\rho_{i-1}}{2}\right)\ .
\end{equation}
For the details  required to implement the method in practice see
 \cite{PhysRevD.85.042003}. The accuracy of this technique is within the calibration error of the
gravitational-wave detector with results similar to the sensitivity estimates
 performed with Monte-Carlo methods, thus we adopt it in  the following numerical 
 optimization scheme \cite{Abadie:2010px,PhysRevD.85.042003}.

\subsection{Sensitivity function}
The signal strength $\hrms$ in the detector, depends on the intrinsic amplitude $h_{0}$, the 
sky-position of the source, the polarization angles and the detector orientation. Averaging isotropically
over the sky-position and polarization yields
\begin{equation}
 \label{eq:8}
\avg{\hrms^{2}}_{\mathrm{sky,pol}} = \frac{2}{25}\hnot^{2}\ .
\end{equation}
Under ideal conditions of data without gaps with duration $\Tobs$ and constant noise floor $S_{n}$
 the accumulated squared $\SNR$ in a semicoherent search is \cite{PrixShaltev2011}:
\begin{equation}
 \label{eq:9}
\hat{\rho}^{2} = 2[1 - \xi(\co{m}+\ic{m})]\frac{2\Ndet}{S_{n}}\hrms^{2}\Tobs\ ,
\end{equation}
where $\xi\in(0,1)$ is the geometrical factor used to estimate the average mismatch 
of the template bank, $\co{m}$ is the mismatch on the coarse grid,
 $\ic{m}$ is the mismatch on the fine grid and $\Ndet$ is the number of detectors.
Typically, the data is made available in the form of $\Nsft$ number of Short Fourier 
Transforms (SFTs) of duration $\Tsft$. To take into account possible noise 
floor fluctuations, the noise floor can be written as \cite{PrixWette2012}
\begin{equation}
\label{eq:10}
\mS^{-1}(f_{0})\equiv\Nsft^{-1}\sum_{n=1}^{\Nsft}S_{n}^{-1}(f_{0})\ ,
\end{equation}
where $S_{n}$ is the per SFT noise Power Spectral Density (PSD) estimated at
frequency $f_{0}$. 
To account for the possible gaps in the data we define the actual amount of available data
\begin{equation}
\label{eq:11}
\Tdata\equiv\Nsft\Tsft\ .
\end{equation}
Using Eqs. \eqref{eq:8}, \eqref{eq:10} and \eqref{eq:11} we can rewrite Eq. \eqref{eq:9} to obtain the
accumulated squared $\SNR$ in a semicoherent search under realistic conditions, namely
\begin{equation}
\label{eq:12}
 \avg{\hat{\rho}^{2}}_{\mathrm{sky, pol}} = \frac{4}{25}[1 - \xi(\co{m}+\ic{m})]h_{0}^{2}\Tdata\mS^{-1}\ .
\end{equation}
To estimate the minimal detectable intrinsic amplitude $\hnot$ at fixed 
false-alarm $\pFA^{*}$ and false-dismissal $\pFD^{*}$ probability we use the
per-segment threshold SNR $\rho^{*}$. With
$\avg{\hat{\rho}^{2}}_{\mathrm{sky, pol}}\equiv\Nseg\rho^{*2}$
 substitution in Eq. \eqref{eq:12} and rearrangement yields
\begin{equation}
 \label{eq:13}
{\hnot}=\frac{5}{2}[1 - \xi(\co{m}+\ic{m})]^{-1/2}\rho^{*}\sqrt{\Nseg}\sqrt{\frac{1}{\mG}}\ ,
\end{equation}
where  
\begin{equation}
\label{eq:14}
\mG \equiv \Tdata\mS^{-1} 
\end{equation}
 is the \textit{goodness} of the data. Eq. \eqref{eq:13} is the  function
 that we need to minimize under the constraint of  fixed computing cost
 $\CC_{0}$ in order to maximize the sensitivity of the search.

\subsection{Computing cost}

The total computing cost $\CCtot(\co{m},\ic{m},\Nseg,\Tseg,\Nsft)$ of the StackSlide method
 is composed by the cost
$\co{\CC}(\co{m},\Tseg,\Nsft)$ to compute the
$\mF$-statistic \cite{jks98:_data,cutler05:_gen_fstat} on the coarse grid and 
the cost
$\ic{\CC}(\ic{m},\Nseg,\Tseg)$ to sum these $\mF$-statistic values 
across all segments on the fine grid, thus
\begin{eqnarray}
\label{eq:15}
 \CCtot(\co{m},\ic{m},\Nseg,\Tseg,\Nsft)&=&\co{\CC}(\co{m},\Tseg,\Nsft)\nonumber\\&+&\ic{\CC}(\ic{m},\Nseg,\Tseg)\ .
\end{eqnarray}
The computing cost of the coherent step using the SFT method
 is
\begin{equation}
 \label{eq:16}
\co{\CC}(\co{m},\Tseg,\Nsft) = \Nsft\co{\Nt}(\co{m},\Tseg)\co{c}_{0}^{SFT}\ ,
\end{equation}
where $\co{\Nt}(\co{m},\Tseg)$ is the number of coarse-grid templates and 
$\co{c}_{0}^{SFT}$ is an implementation- and hardware-dependent fundamental 
computing cost. Similarly, the incoherent computing cost is
\begin{equation}
 \label{eq:17}
\ic{\CC}(\ic{m},\Nseg,\Tseg) = \Nseg\ic{\Nt}(\ic{m},\Tseg,\Nseg)\ic{c}_{0}\ ,
\end{equation}
where $\ic{\Nt}(\ic{m},\Tseg,\Nseg)$ is the number of fine-grid templates
and  $\ic{c}_{0}$ is the fundamental cost of adding $\mF$-statistic values.
 
\subsection{Templates counting}
The general expression for the number of templates required to cover some
parameter space $\dopS$ is 
\begin{equation}
  \label{eq:18}
  \Nt = \thick_n \, \mismax^{-n/2}\, \vol_n\,,\quad\mbox{with}\quad
  \vol_n \equiv \int_{\dopT_n}\mathrm{d}^{n}\dop\,\sqrt{\det g}\,,
\end{equation}
where $\thick_n$ is the normalized thickness of the search grid, $m$ is the
maximal-mismatch, $\det g$ is the determinant of the parameter-space metric
$g_{ij}$ and $\vol_n$ is the metric volume of the $n$-dimensional space of the
template bank. For hyper-cubic lattice the normalized thickness is $\thick_{\mathbb{Z}_n} =
n^{n/2}\,2^{-n}$, while for an $A^{*}$ lattice it is 
$\thick_{A^{*}_{n}}=\sqrt{n+1}\left\{\frac{n(n+2)}{12(n+1)}\right\}^{n/2}$ \cite{2007arXiv0707.0428P}.
 The choice of the dimensionality of the template bank is
subject to the maximization of the number of templates, namely
\begin{equation}
  \label{eq:19}
  \co{\Nt}_{\nc} = \max_{n} \co{\Nt}_n\,,\quad\mbox{and}\quad
  \ic{\Nt}_{\nf} = \max_{n} \ic{\Nt}_n\,.
\end{equation}
In Ref. \cite{PrixShaltev2011} we used the factorization of the semicoherent
 metric volume
\begin{equation}
  \label{eq:20}
  \ic{\vol}_\nf(\Nseg,\Tseg) = \refac_{\nf}(\Nseg)\,\co{\vol}_\nf(\Tseg)\,,
\end{equation}
to derive the general power-law computing-cost model, as in the gap-less data
 case, the refinement factor $\refac_{\nf}(\Nseg)$ is only a function of the number of segments $\Nseg$.
 However using real data introduces an additional dependency on the time span of
 the search through the actual position of the segments in time. For details see, e.g., \cite{Pletsch:2010a}.
 We aim to model the real conditions as closely as possible, therefore
in the numerical optimization we directly compute the semicoherent metric
\begin{equation}
\label{eq:21}
 \ic{g}(t_{\mathrm{ref}}) = \frac{1}{\Nseg}\sum_{i=1}^{\Nseg}\co{g_i}(t_{i},\Tseg,t_{\mathrm{ref}})\ ,
\end{equation}
where $\co{g}_{i}(t_{i},\Tseg,t_{\mathrm{ref}})$ is the coherent metric of segment $i$
at fixed reference time $t_{\mathrm{ref}}$. 
To compute the coherent metric $\co{g}_{i}(t_{i},\Tseg,t_{\mathrm{ref}})$ in this work
we use the analytical expressions found in Ref. \cite{Pletsch:2010a}.

\subsection{The choice of spindown parameter space}
The choice of the dimensionality of the template bank through 
Eqs. \eqref{eq:19} is possible only for a simple rectangular geometry
of the parameter space.

Using the spindown age of a potential source
\cite{Brady:1998nj}:
\begin{equation}
 \label{eq:22}
 \tau = f/\abs{\dot{f}}\ .
\end{equation}
the spindown search band in dimension $k$ is 
\begin{equation}
\label{eq:23}
 \abs{f^{(k)}}\le k!\frac{f}{\tau^{k}}\  .
\end{equation}
This however means that the spindown band is frequency dependent, which
may be impractical.  Therefore if we keep the number of templates
in the spindown space constant by fixing a minimal detectable
spindown age $\tau_{0}$ at some frequency $f_{0}$, the detectable spindown
age at frequency $f$ yields
\begin{equation}
 \label{eq:24}
 \tau(f) = \tau_{0}f/f_{0}\ .
\end{equation}
This would define the simplest possible parameter-space volume for
 optimization, namely a ``box''. 

A more complicated triangular parameter-space shape has been discussed 
in Ref. \cite{2008CQGra..25w5011W} and used in the search for gravitational waves
from the supernova remnant Cassiopeia-A  \cite{2010ApJ...722.1504A}.  
The parameters of a search over such space are difficult to optimize
as the spindown order may vary even in infinitesimally small slices of
 a frequency band. In the present work we neglect
 this fact in order to compare the outcome of the optimization with 
 previously obtained results.

\section{Numerical optimization procedure for a semicoherent StackSlide search with a fixed frequency band}
\label{sec:3}
\begin{figure*}[ht]
\subfigure[Greedy data selection for requested $\Nseg=10$ segments with
 duration $\Tseg=3$ time units. Note how segments 8, 9 and 10 overlap in time
 with their neighboring segments, however they do not share data. This is
 depicted by the black fill of the segments.]{\includegraphics[width=0.9\linewidth]{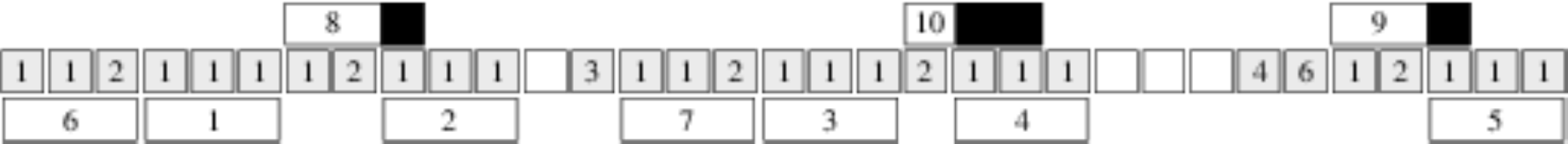}\label{fig:dselect}}
 \subfigure[Compact data selection for requested $\Nseg=2$ segments
 with duration $\Tseg=3$ time units. We show only the best segment
 combination. For comparison with the greedy data
 selection method we denoted the selected segments of the greedy
 procedure with G. The PSD of the SFTs has been chosen such that the best
segment combination of the compact method has much shorter time
 span.]{\includegraphics[width=0.9\linewidth]{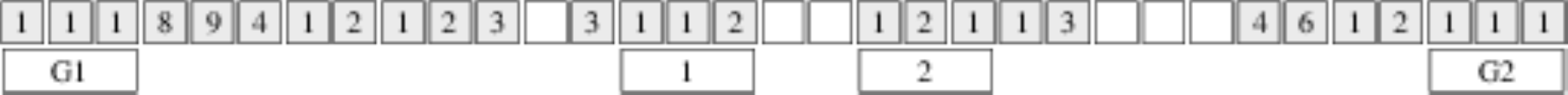}\label{fig:cselect}}
\caption[Data selection]{Schematic representation of (a) greedy  and (b) compact
data selection. The middle line of squares in gray are the
 available SFTs of unit time, where the number inside the square denotes
 the PSD. The white squares without a number are gaps in the data. The
 selected segments are the white rectangles, where the number inside is the
 number of the segment.}
\end{figure*}
In this section we consider the practical implementation of a numerical optimization
procedure to find optimal search parameters for a semicoherent StackSlide search.

\subsection{Definition of the optimization problem}

To maximize the sensitivity of the search, i.e., to minimize the 
measurable intrinsic amplitude, we need to minimize the function  
given in Eq. \eqref{eq:13}, namely:
\begin{equation}
 \label{eq:25}
\hnot(\co{m},\ic{m},\Nseg,\mG) = \frac{5}{2}[1 -
\xi(\co{m}+\ic{m})]^{-1/2}\rho^{*}\sqrt{\Nseg}\sqrt{\frac{1}{\mG}}
\end{equation}
under the constraints
\begin{eqnarray}
 \label{eq:26}
 \CCtot&\le&C_{0}\ ,\nonumber\\
 0<&\co{m}&<1\ ,\nonumber\\
 0<&\ic{m}&<1\ ,\nonumber\\
 \co{\Nt}&\le&\ic{\Nt}\ ,
 \end{eqnarray}
where $\CC_{0}$ is a given maximal available computing cost. 
Note that in practice the cost constraint is difficult to be fulfilled 
as an equality, however a reasonable algorithm should find a solution for which $\CCtot$ 
becomes approximately $\CC_{0}$. 
In order to minimize $\hnot$ we need a data selection procedure which maximizes
the goodness of the data $\mG$. Through the data selection procedure we can 
transform the implicit dependency of Eq. \eqref{eq:25} on the segment duration 
$\Tseg$ into an explicit dependency and minimize the 4D optimization problem 
$\hnot(\co{m},\ic{m},\Nseg,\Tseg)$.

\subsection{Data selection}

For a given amount of data and requested number of segments of some duration,  we 
need an algorithm to select the data which maximizes the goodness $\mG$, i.e., 
pick as much data $\Tdata$ as possible of lowest noise level $\mS$ as possible
\footnote{For a summary of data selection methods used in past searches 
for continuous gravitational waves see \cite{2015PhRvD..91f4007B}.}.  
This would require computation of all possible segment combinations, their ranking 
by the goodness $\mG$ and selection of the first segment realization which 
satisfies the computing cost constraint. For the simple case of picking $k$ 
non-overlapping segments out of $n$ possible, without replacement and ordering, 
the number of combinations is given by the well known binomial coefficient formula
\begin{equation}
\label{eq:27}
 \left(\begin{tabular}{c}
        $n$\\$k$
       \end{tabular}
\right) = \frac{n!}{k!(n-k)!}\,.
\end{equation}
For example, having 200 days of available data, choosing 100 segments of 1 day
duration, yields roughly $9\times10^{58}$ possible combinations. Clearly such 
data selection procedure is not well suited for practical implementation
\footnote{Similarly we can argue that per SFT data optimization procedure
is completely computationally unfeasible.}. Therefore
we consider three different suboptimal, but computationally feasible, alternative 
methods for data selection. Namely the two extremes,  a \textit{greedy} method and a
 \textit{compact} method, and a third procedure \textit{greedy-compact}, which is 
placed somewhere in between, in terms of the total observation time.
 
\subsubsection{A greedy method}
For requested $\Nseg$  number of segments with 
duration $\Tseg$ and given set of SFTs with duration $\Tsft$,
 which are ordered in time by increasing timestamps $t_{j}$,
 within a greedy data selection we always pick the segments with the maximal
 goodness. The steps of the algorithm are (pseudo code is given in 
 Alg.~\ref{alg:greedy-data-selection})
\begin{enumerate}
 \item For each timestamp $t_{j}$ find all SFTs in the time interval $[t_{j},t_{j}+ \Tseg]$
 and compute $\mG_{j}$.
 \item Select the segment starting from $t_{j}$ such that $\mG_{j}$ is maximal
 and remove the timestamps of  the SFTs, which belong to the selected segment.
 \item Repeat steps 1. and 2. until $\Nseg$ segments are selected or there is no 
more data left.
\end{enumerate}
An example of data selection is schematically presented in Fig.~\ref{fig:dselect},
where we select $\Nseg=10$ segments of duration $\Tseg=3$ time units out of
data set with $\Nsft = 33$ SFTs of unit time distributed in $\Tobs=37$ time units.
Three of the selected segments overlap in time with their neighbors, however
 these overlapping segments do not share data. Such partial segments
 suggest grid construction based on the actual length of the segments instead of
the maximal length $\Tseg$, which would reduce the coherent part of the total
 computing cost in these cases. However this would complicate the combination
of the $\mF$-statistic values in the semicoherent step of the StackSlide search,
thus the overall effect remains unclear. Therefore we stick to a constant grid for
every segment. While the above criticism also holds for the other two data 
selection methods proposed in this subsection, we should note a specific weak
point of the greedy algorithm. Namely, depending on the quality of the data this 
method tends to maximize the total observation time of the generated segment 
list, for example when the data is of low noise level at the beginning and at the end,
but contains disturbences or gaps in the middle. In such cases an alternative 
method yielding more compact segment list may lead to higher search sensitivity.

Before proceeding with the explanation of the compact data selection, we note,
 that an equivalent  procedure of the greedy method has been used to select data
 for  the recent all-sky Einstein@Home searches for continuous gravitational 
 waves in LIGO S6 data \cite{EAH:Misc}.

\subsubsection{A compact method}

For requested $\Nseg$ number of segments with 
duration $\Tseg$ and given set of SFTs with duration $\Tsft$,
 which are ordered in time by increasing timestamps $t_{j}$,
 the compact data selection aims to maximize the goodness while keeping the
 total timespan of the data close to the minimum $\Nseg\Tseg$.
 The algorithm consist in the following steps (pseudo code is given in 
 Alg.~\ref{alg:compact-data-selection})
\begin{enumerate}
\item Set the start time of the first segment to $t_{s} = t_{j}$.
\item Find all SFTs in the time interval $[t_{s},t_{s} + \Tseg]$.
\item Update $t_{s}$ with the first $t_{j} \ge t_{s} + \Tseg$.
\item Repeat steps 2. and 3. until N segments are selected or there is
no more data left.
\item Compute $\mG_{j}$.
\item Repeat steps 1. to 5. for all timestamps $t_{j}$.
\item Sort all found combinations by decreasing $\mG$.
\end{enumerate}
Using the compact method of data selection, we obtain a list of 
possible segment combinations. Then we use the first combination, 
which satisfies the computing cost constraint. 
An example of compact data selection is schematically presented in 
Fig.~\ref{fig:cselect}, where we select $\Nseg = 2$ segments of duration 
$\Tseg = 3$ time units out of
 data set with $\Nsft = 33$ SFTs of unit time distributed in $T = 37$ time 
units. For comparison with the greedy method we also 
show the outcome of the greedy procedure. This example is specially
constructed to stress the difference between the compact and the greedy 
method in terms of the time spanned by the data. 

Due to the complexity of the compact method, but also to fill the gap to the 
greedy data selection algorithm, we also consider a third method, namely
a greedy-compact algorithm.

\subsubsection{A greedy-compact method}

This method lies between the greedy and the compact methods in terms of the 
total span of the selected data. We achieve that by maximization of the
sensitivity per cost ratio $\hnot^{-1}/\CC$. Using Eq. \eqref{eq:25} this is 
equivalent to maximization of the ratio $\mG/\CC^{2}$. The steps of the 
algorithm are (see Alg.~\ref{alg:greedy-compact-data-selection} for pseudo code)
 \begin{enumerate}
 \item For each timestamp $t_{j}$ find all SFTs in the time interval
 $[t_{j},t_{j}+ \Tseg]$
 and compute $\mG_{j}/\CC_{j}^{2}$, where $\CC_{j}$ is the computing cost
resulting from using this particular segment list and $\CC_{1}=1$.
 \item Select the segment starting from $t_{j}$ such that $\mG_{j}/\CC_{j}^{2}$
 is maximal and remove the timestamps of  the SFTs, which belong to the selected
 segment.
 \item Repeat steps 1. and 2. until $\Nseg$ segments are selected or there is no 
more data left.
\end{enumerate}

In the next section we compare the results of the optimization procedure 
using the three different data selection methods.

\section{Examples of practical application}
\label{sec:4} 

In the absence of direct detection limits on the 
intrinsic gravitational-wave amplitude have been set 
for example, in Refs.
\cite{2007arXiv0708.3818L,Abbott:2009nc,Aasi:2012fw,2010ApJ...713..671A,Aasi:2013jya,Aasi:2014ksa}
The intrinsic gravitational-wave amplitude and with this the 
probability of detection depends on unknown priors, in particular 
on the population and ellipticity of the continuous waves emitters. 
Thorough population studies and prospects for detection with
the initial and Advanced LIGO \cite{TheLIGOScientific:2014jea} 
detectors can be found in Refs. \cite{Palomba:2005na,Knispel:2008ue,Wade:2012qc}.
In this work it is convenient to use as a figure of merit the sensitivity depth
\begin{equation}
\label{eq:28}
 \sensdep = \frac{\sqrt{\mathcal{S}}}{\hnot}\,,
\end{equation}
a quantity introduced in Ref. \cite{2015PhRvD..91f4007B}. With this we use 
for comparison the analytical solution found in the ``Directed search for Cassiopeia-A`` 
example in Ref. \cite{PrixShaltev2011}, therefore
we use the same search volume enclosed in the frequency band
 $f\in[100,300]\,\Hz$ with spindown ranges corresponding to a spindown age
$\tau_{\min}=300\,\years$. The computing cost constraint is
$\CC_{0}\approx472$ 
days on a single computing core, where the fundamental computing
 constants  are
\begin{equation}
 \label{eq:29}
\co{c}_{0}^{\SFT} = 7\times10^{-8}\ \secs,\ \ \ \ \ic{c}_{0}=6\times10^{-9}\
\secs\ .
\end{equation}
In the following we assume an $A^{*}$ search
 grid, for which $\xi\approx0.5$. We fix the false-alarm $\pFA=1\times10^{-10}$ and 
false-dismissal probability $\pFD=0.1$.
The weakest detectable signal, as estimated for
 $\Tdata = 0.7\times2\times12\,\Days$, $\pacfactor = 0.5$ and $\co{m}=0.2$ yields
 sensitivity depth of
\begin{equation}
 \label{eq:30}
\left.\sensdep\right|_{\mathrm{opt}}\approx54.4\,\Hz^{-1/2}\,,
\end{equation}
which in the KWS approximation yields
\begin{equation}
 \label{eq:31}
\left.\sensdep^{\KWS}\right|_{\mathrm{opt}}\approx36.9\,\Hz^{-1/2}\,.
\end{equation}

We perform the numerical optimization with the NOMAD
 \cite{LeDigabel2011A909} implementation of  a Mesh Adaptive Direct Search
 (MADS)
 \cite{Audet04meshadaptive,AbramsonADD09,LeDigabel2011}
 algorithm for constrained derivative-free optimization. For each of the
following
 examples we run the procedure 50 times from a common initial starting
 point:
\begin{eqnarray*}
 \Nseg_{0} = 200\,,\quad\Tseg_{0} = 1\,\Days\,\\\nonumber
 \co{m}_{0} = 0.5\,,\quad \ic{m}_{0}=0.5,\\\nonumber
\end{eqnarray*}
 while we use different mesh coarsening and mesh update basis
 parameters \footnote{These are internal parameters for the MADS algorithm
 controlling the evolution of the mesh, if a better solution than the
 current one is found.}. We use these multiple runs of the optimization
 effectively to escape local extremes. Note that in Figures \ref{fig:DC100} to 
 \ref{fig:S5EAH} we plot the best solution from each of these 50 runs,
 however on average approximately $17\times10^3$ points have been evaluated 
 per data selection algorithm for each of the studied cases.

 In all cases we use the three different data selection
algorithms.

\subsection{Directed search using simulated data}

\subsubsection*{Gapless data with constant noise floor}

We first consider  optimization using simulated data from 2 detectors spanning
365 days, without gaps, and with a constant noise floor $\sqrt{\Sn}=1\ \Hz^{-1/2}$. 
Using the analytical optimization method discussed in \cite{PrixShaltev2011} and
the WSG approximation to obtain optimal parameters, the sensitivity depth
of the search expressed in the KWS approximation is
\begin{equation}
\label{eq:32}
 \left.\sensdep^{\KWS}\right|_{\mathrm{opt}}\approx78.6\,\Hz^{-1/2}\ .
\end{equation}
The  optimal maximal mismatch on the coarse and fine grid is
\begin{equation}
 \label{eq:33}
\begin{split}
\opt{\co{m}}=0.16\,,&\quad
\opt{\ic{m}}=0.24\ ,
\end{split}
\end{equation}
and the optimal number of segments $\opt{\Nseg}$, segment duration $\opt{\Tseg}$
and total observation
time $\opt{T}$ are
\begin{equation}
 \label{eq:34}
  \begin{split}
    \opt{\Nseg} = 76.5\,,&\quad
    \opt{\Tseg} \approx 2.0\,\Days\,,\\
    \opt{\Tobs} &\approx 155.5\,\Days\,.
  \end{split}
\end{equation}

\begin{figure*}
\centering
  \subfigure[ ]{\includegraphics[width=0.48\linewidth]{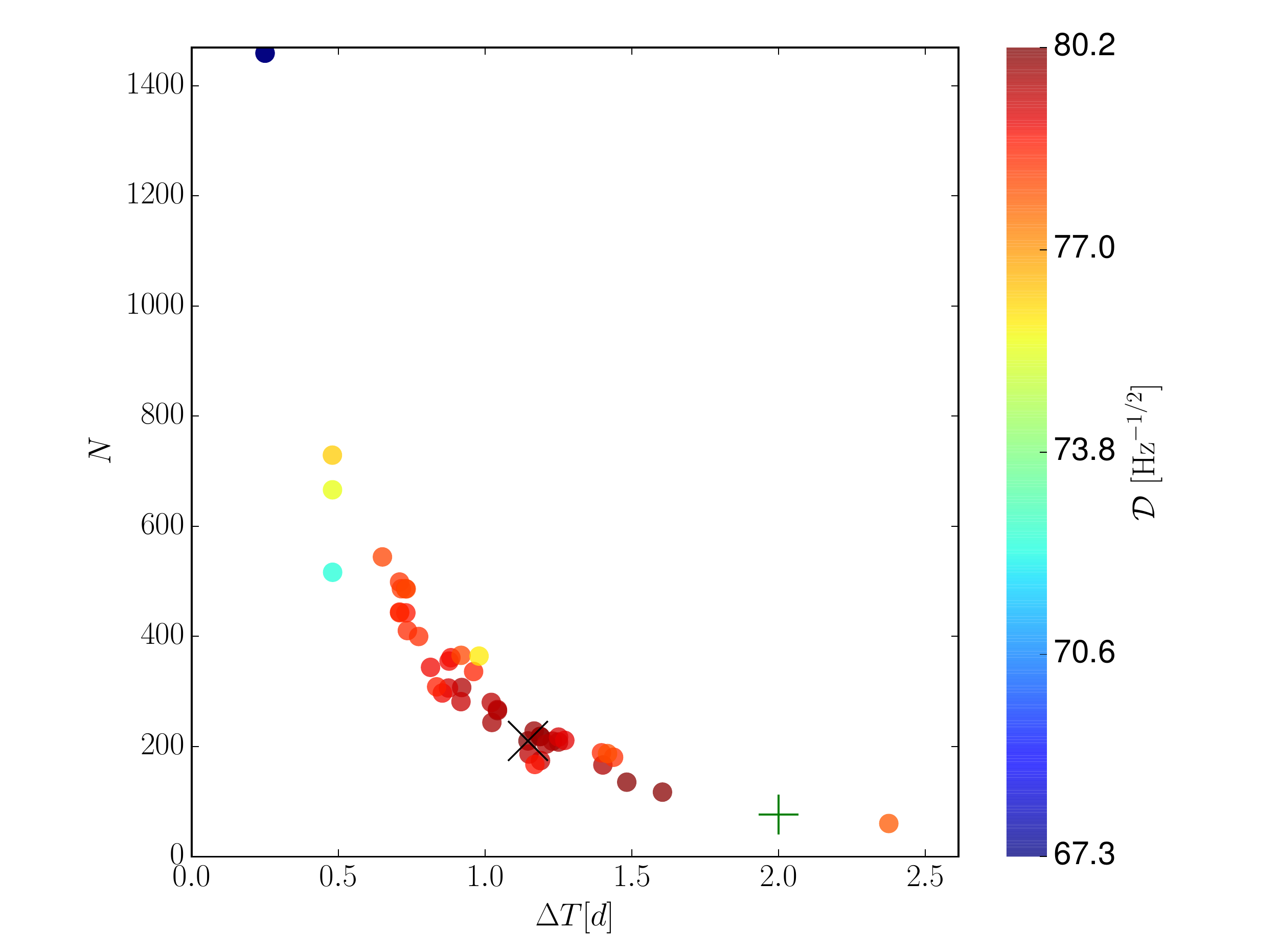}\label{fig:DC100DS1_hnot}}
  \subfigure[ ]{\includegraphics[width=0.48\linewidth]{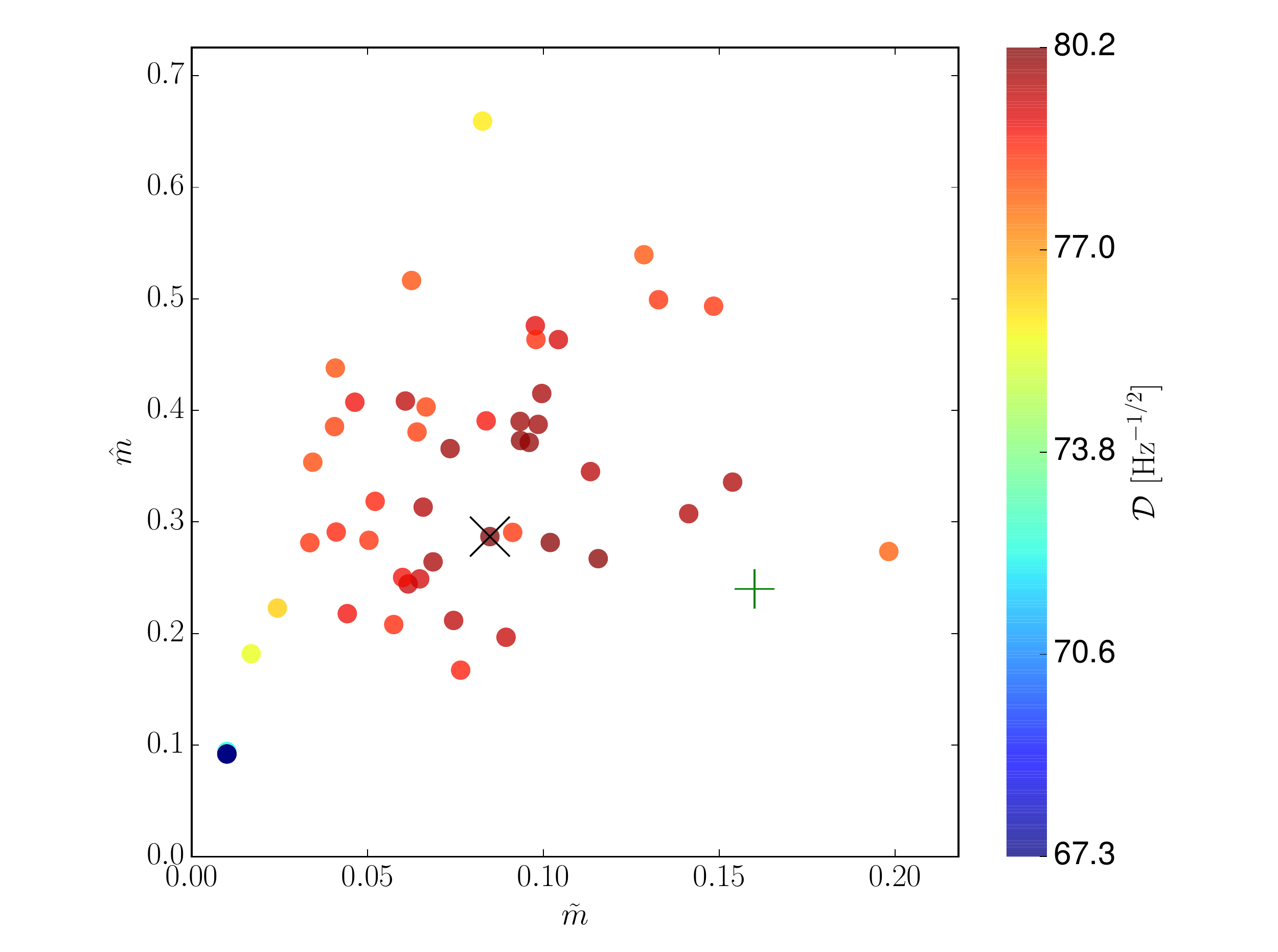}\label{fig:DC100DS1_m}}
\quad
  \subfigure[ ]{\includegraphics[width=0.48\linewidth]{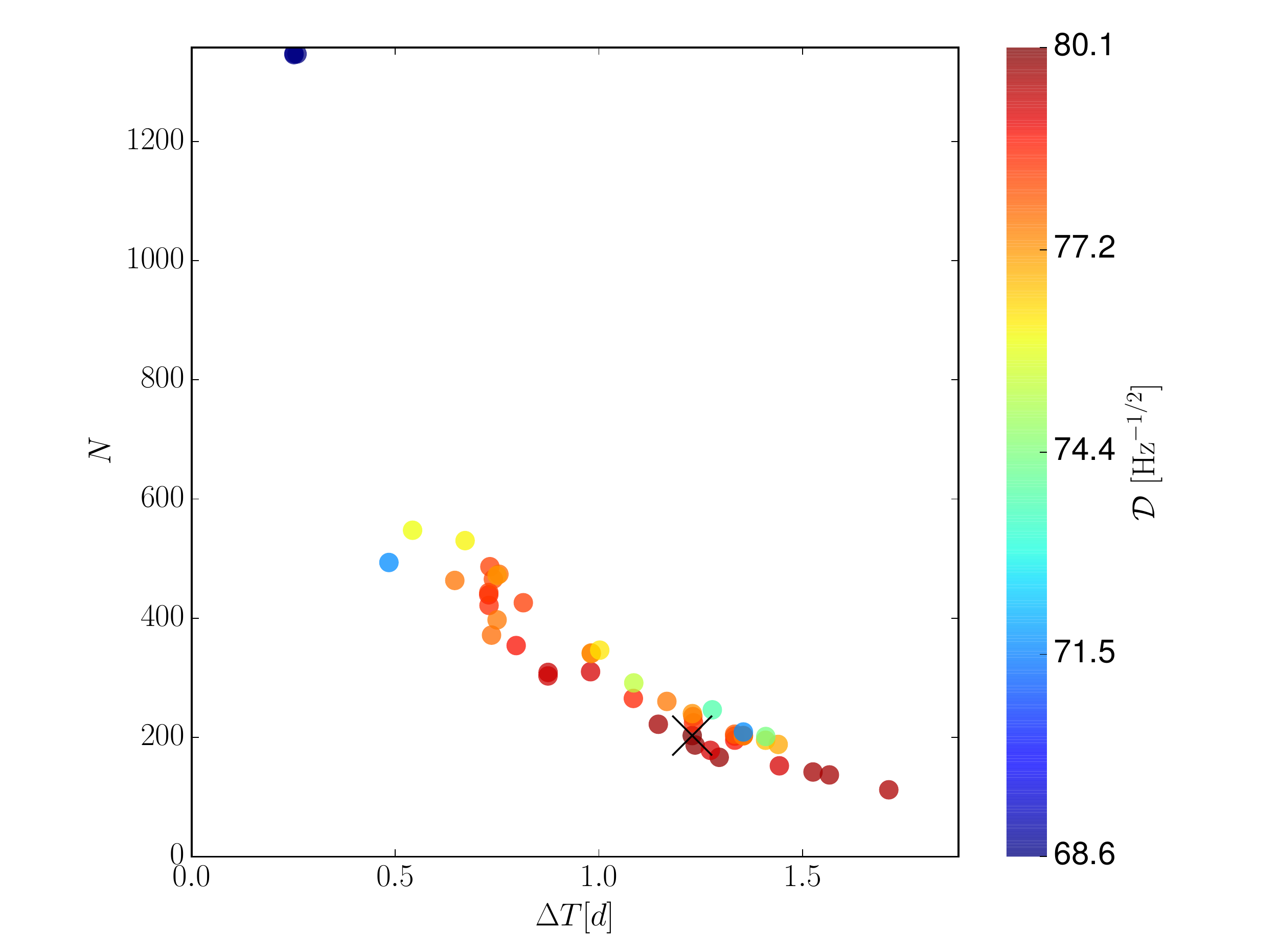}\label{fig:DC100DS0_hnot}}
  \subfigure[ ]{\includegraphics[width=0.48\linewidth]{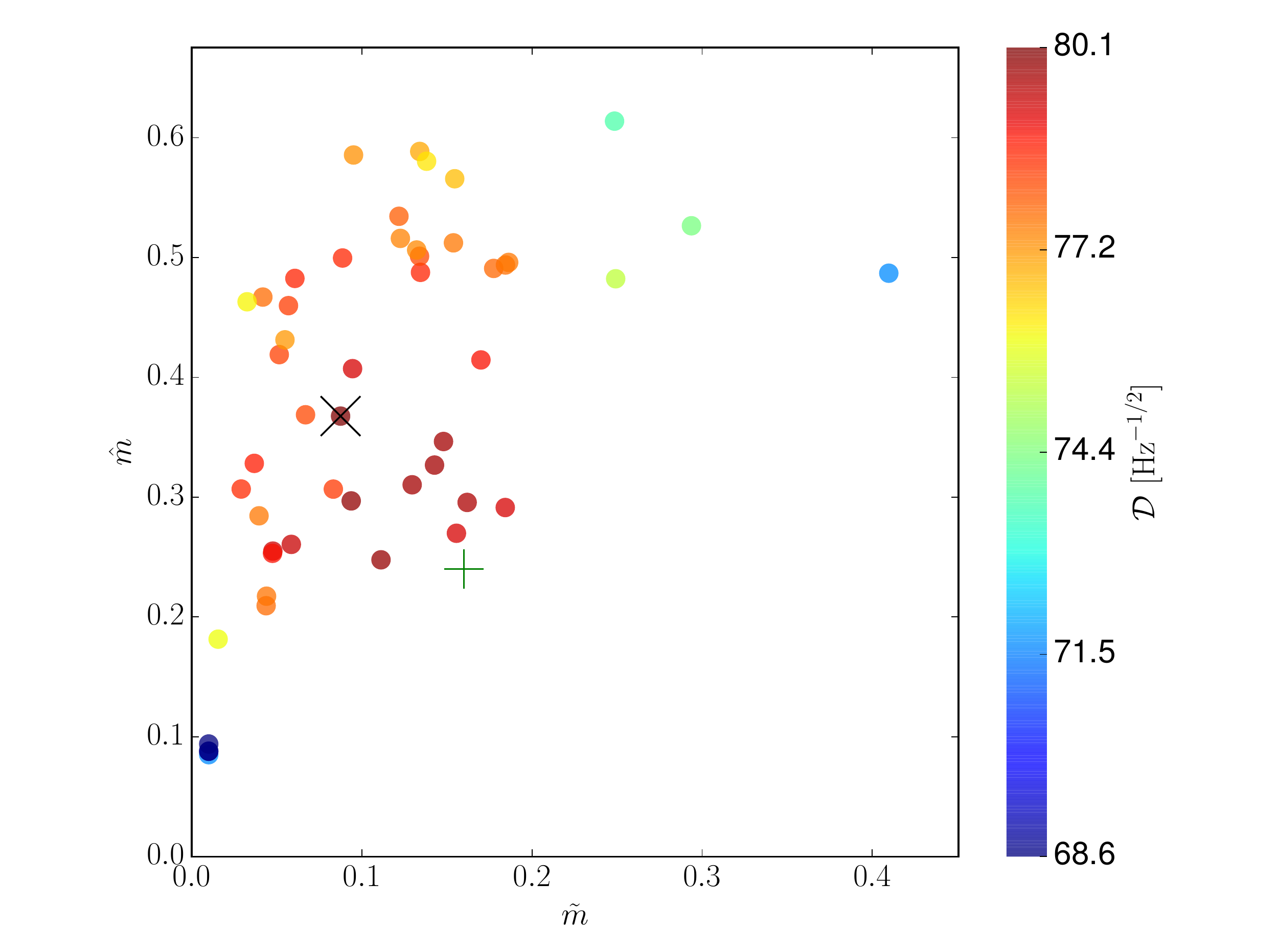}\label{fig:DC100DS0_m}}
\quad
  \subfigure[ ]{\includegraphics[width=0.48\linewidth]{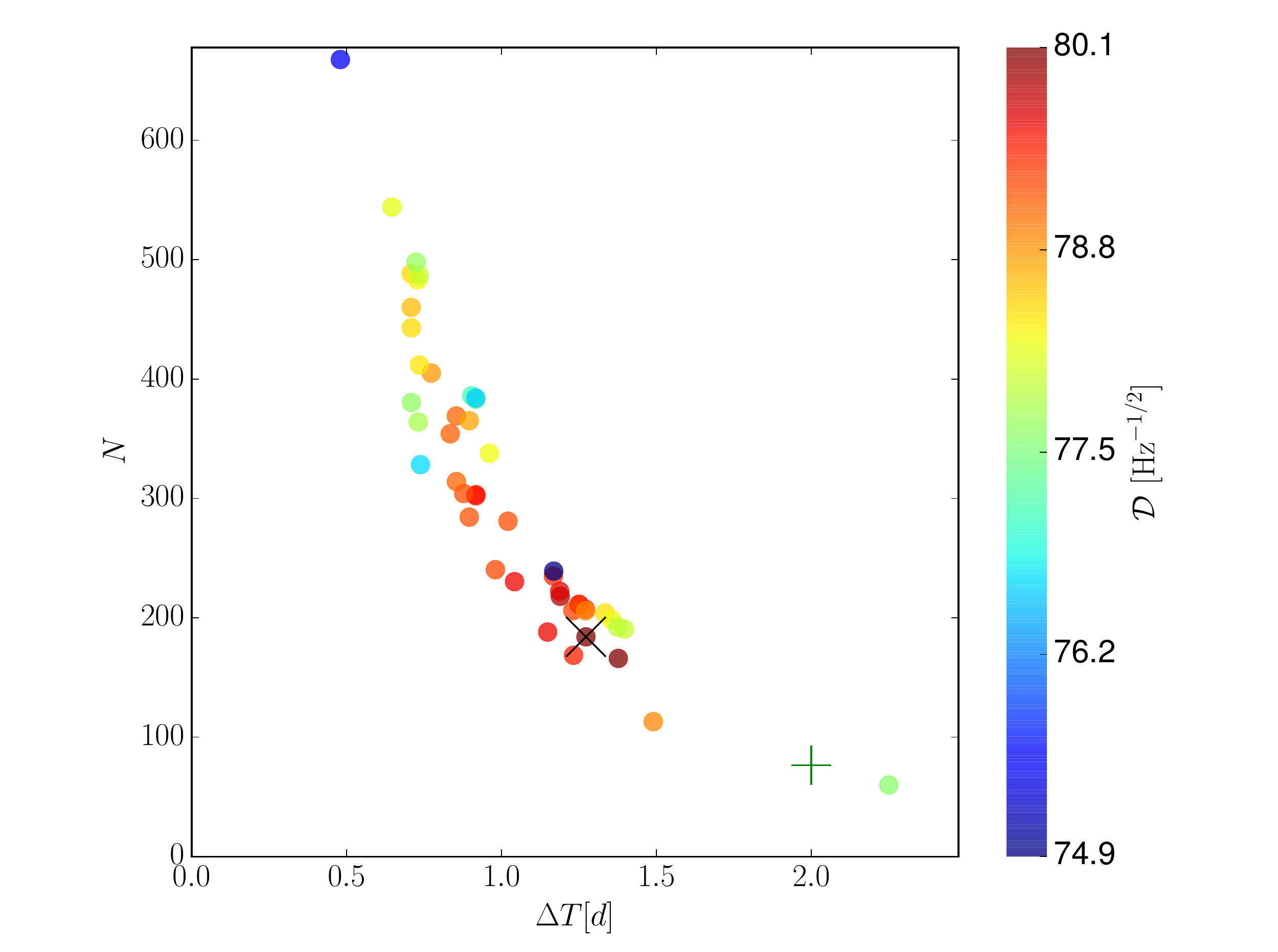}\label{fig:DC100DS2_hnot}}
  \subfigure[ ]{\includegraphics[width=0.48\linewidth]{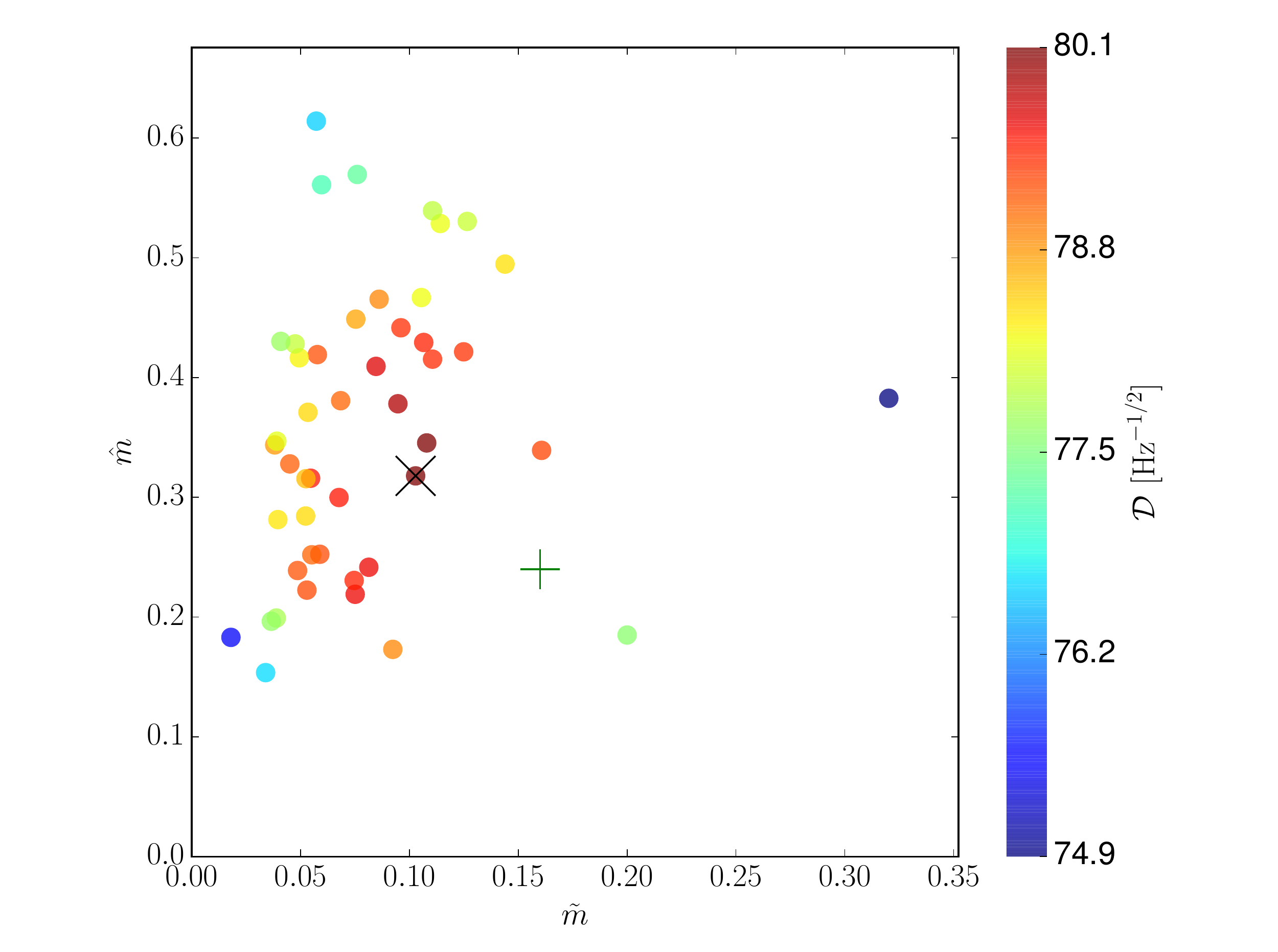}\label{fig:DC100DS2_m}}

\caption[Numerical optimization using ideal data.]{Semicoherent search optimization  for data from 2 detectors, without gaps and of
constant noise floor $\sqrt{\Sn}=1\ \Hz^{-1/2}$. The cost constraint is
$\CC_{0}=472.0\ \days$. The best numerical
solution is denoted with $\times$ and the optimal analytical solution
 with $+$.  The panels on the left side show optimal segment duration
$\Tseg$ and number of segments $\Nseg$, and the panels on the right
side show optimal coherent $\co{m}$ and semicoherent $\ic{m}$
mismatch.  Panels (a) and (b) are obtained using greedy data selection,
the most sensitive optimal solution is
 \input{DC100DS1_optimal.tex}. Panels (c) and (d) are
 obtained using compact data selection, the most sensitive optimal
 solution is \input{DC100DS0_optimal.tex}. Panels (e) and (f)
are obtained using greedy-compact data selection, the most sensitive
optimal solution is \input{DC100DS2_optimal.tex}}
\label{fig:DC100}
\end{figure*}

The results of the numerical optimization performed with the 
three different data selection algorithms are plotted in
 Fig.~\ref{fig:DC100}. In Table \ref{tab:caseA} we summarize the
 found optimal solutions. The three data selection methods in this
case lead to equal sensitivity, which is expected, given that the data
is ideal, i.e., of constant noise floor and without gaps. The small
 deviation in the optimal found parameters is due to the usage
 of the numerical optimization method. 

 Using the more accurate KWS method, the gain in sensitivity of this toy semicoherent 
 search compared to the original fully coherent search is approximately $2.2\ $.

\begin{table}[htb]
\begin{tabular}{|c|c|c|c|}\hline
 & greedy & compact & greedy-compact \\\hline\hline
$\sensdep\,[\Hz^{-1/2}]$ & $80.2$ & $80.1$ & $80.1$\\
$T\,[\days]$ & 245 & 253.8 & 239.3 \\
$\Nseg$ & 210 & 203 & 184.1 \\
$\Tseg\,[\days]$ & 1.1 & 1.2 & 1.3 \\
$\co{m}$ & 0.08 & 0.09 & 0.10 \\
$\ic{m}$ & 0.29& 0.37 & 0.32 \\\hline
\end{tabular}

\caption{Optimal solution using greedy, compact and
 greedy-compact data selection applied to data from 2 detectors,
without gaps and of constant noise floor $\sqrt{\Sn}=1\ \Hz^{-1/2}$.}
 \label{tab:caseA}
 \end{table}

\subsubsection*{Data with gaps and constant noise floor}

\begin{figure*}[htbp]
  \subfigure[ ]{\includegraphics[width=0.49\linewidth]{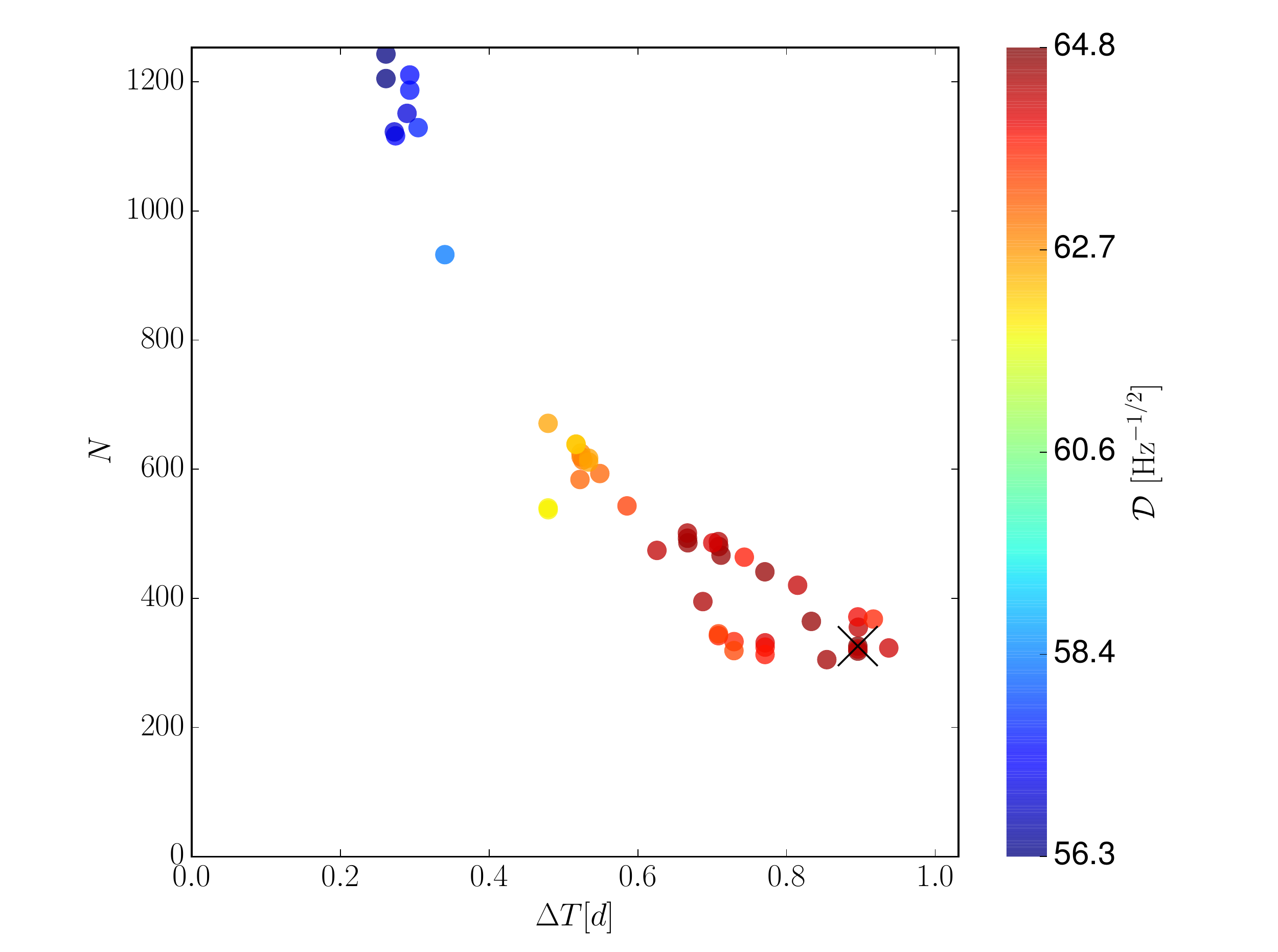}
\label{fig:DC70DS1_hnot}}
 \subfigure[ ]{\includegraphics[width=0.49\linewidth]{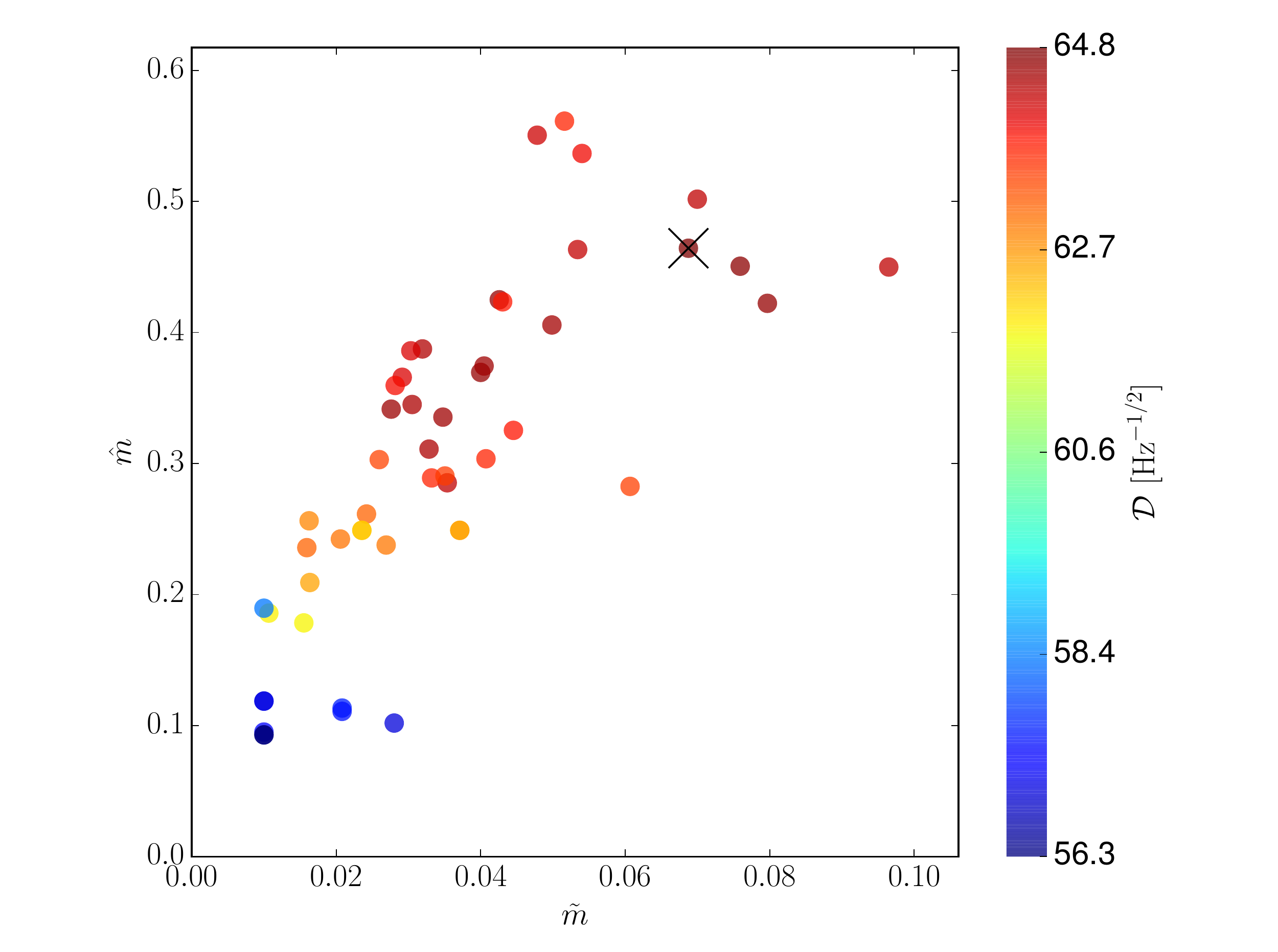}\label{
fig:DC70DS1_m}}
\quad
  \subfigure[ ]{\includegraphics[width=0.49\linewidth]{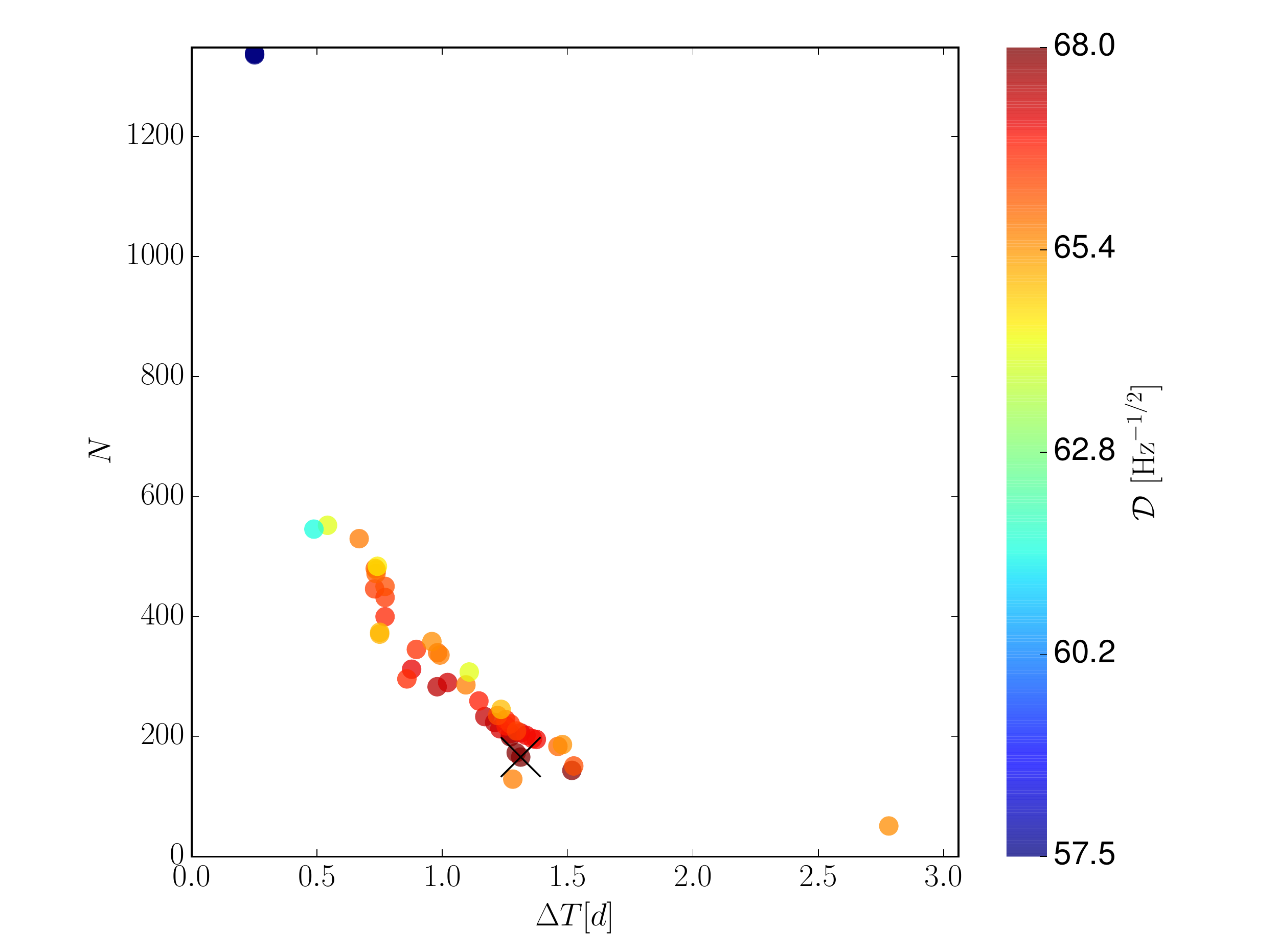}
\label{fig:DC70DS0_hnot}}
 \subfigure[ ]{\includegraphics[width=0.49\linewidth]{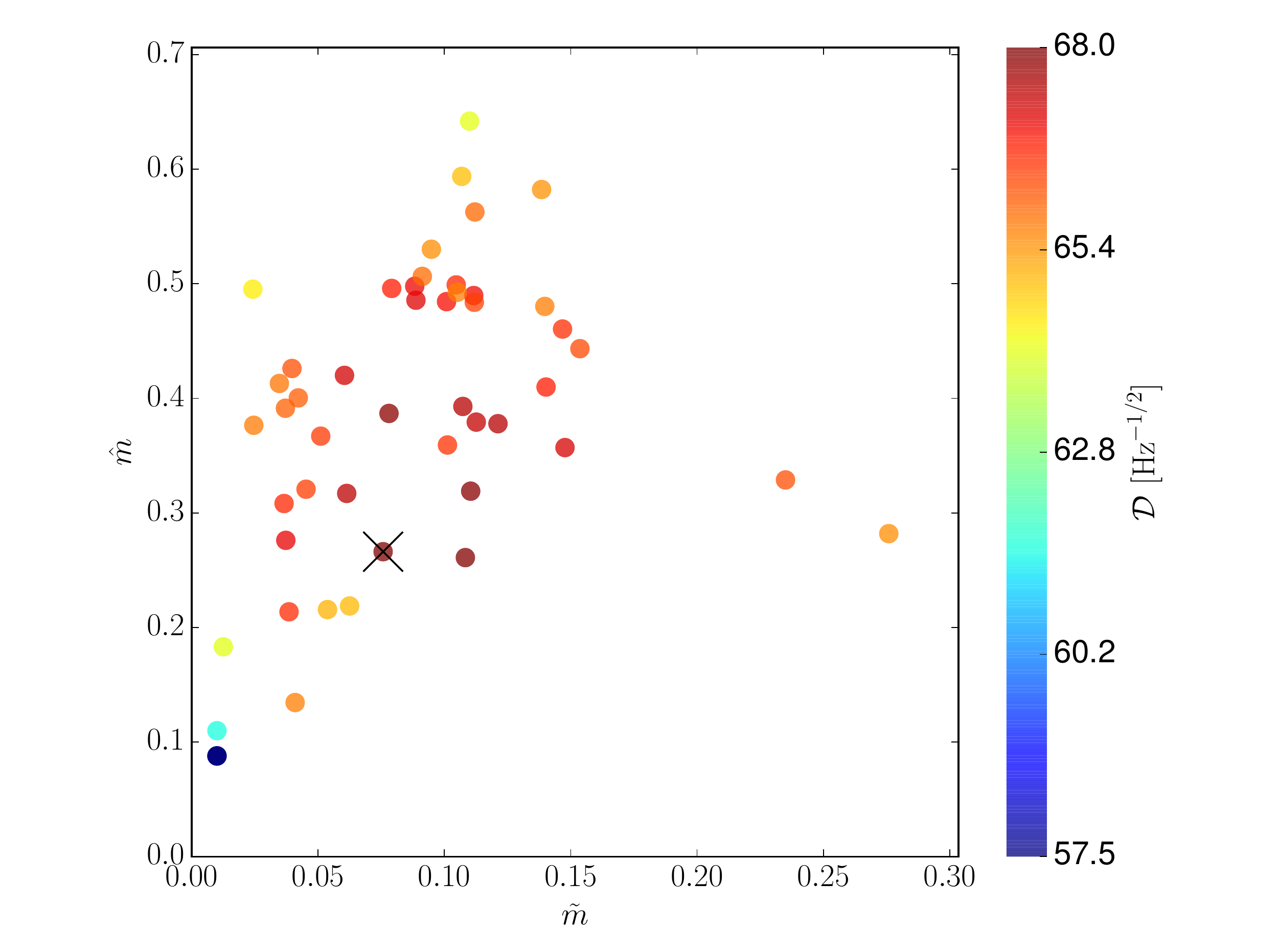}\label{
fig:DC70DS0_m}}
\quad
  \subfigure[ ]{\includegraphics[width=0.49\linewidth]{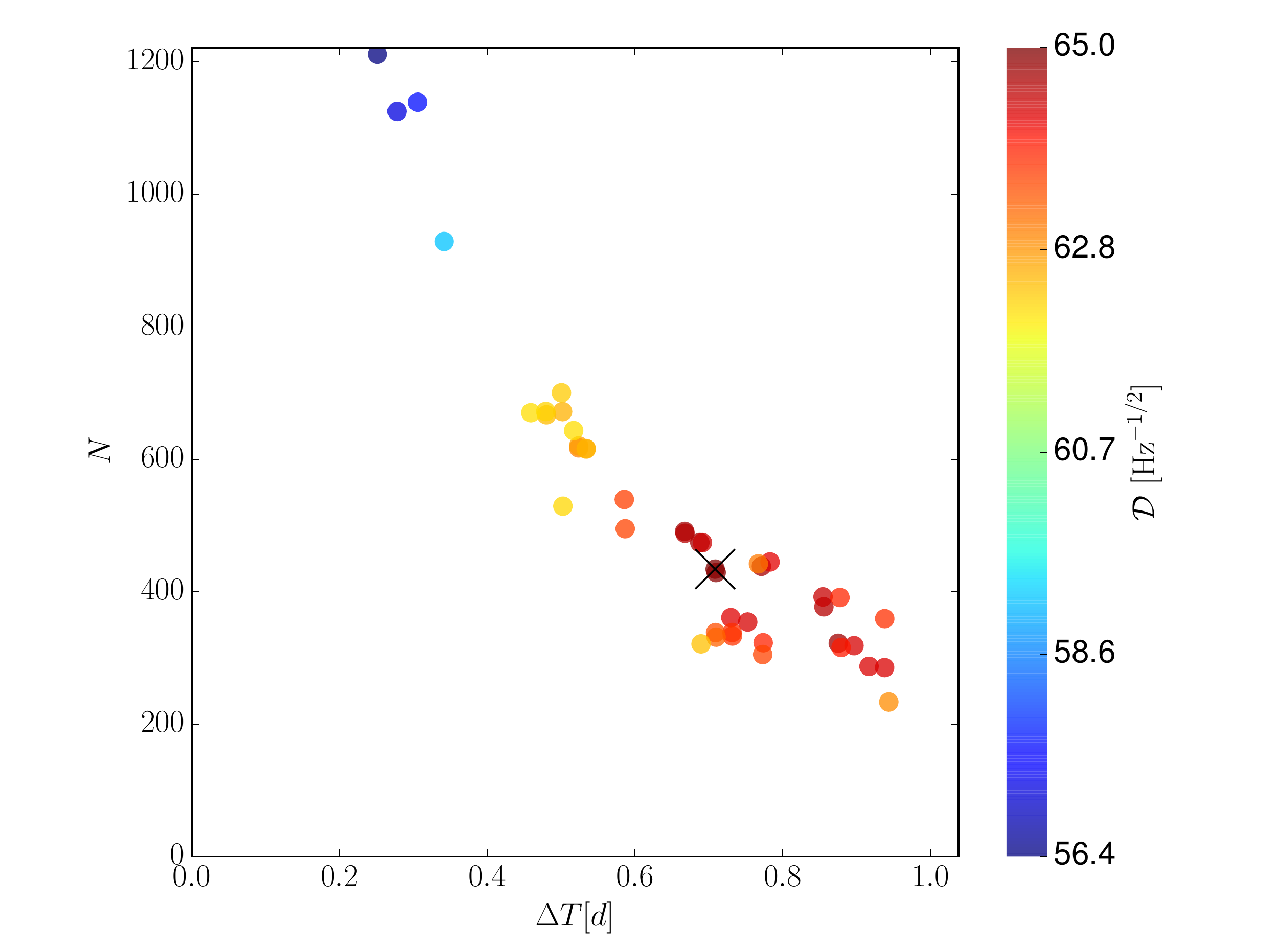}
\label{fig:DC70DS2_hnot}}
 \subfigure[ ]{\includegraphics[width=0.49\linewidth]{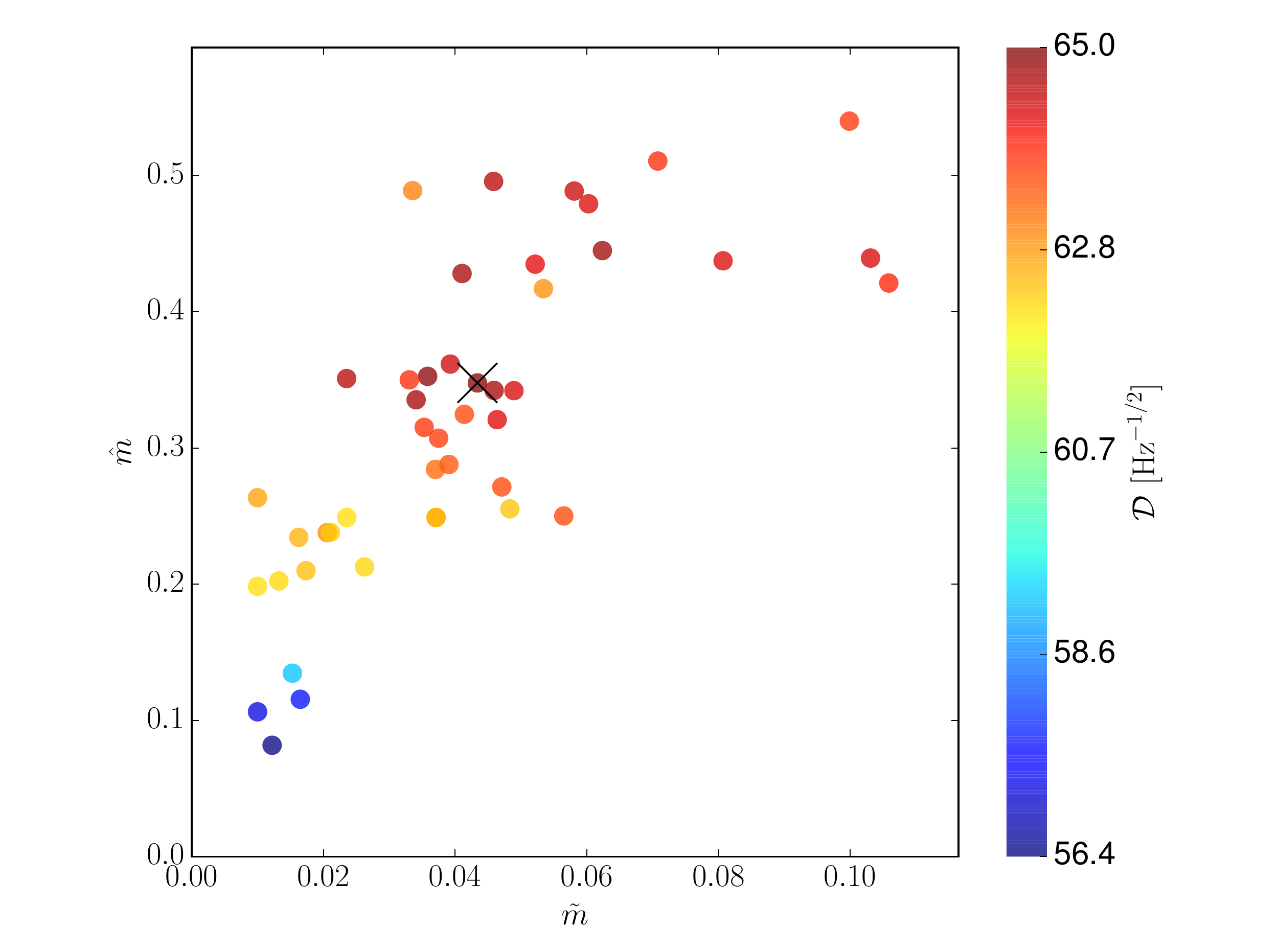}\label{
fig:DC70DS2_m}}
\caption[Numerical optimization using data with constant noise-floor and gaps.]{Semicoherent
 search optimization using data from 2 detectors 
with duty cycle of 70 \%  and of constant noise floor $\sqrt{\Sn}=1\ \Hz^{-1/2}$. 
The cost constraint is
$\CC_{0}=472.0\ \days$. The best numerical
solution is denoted with $\times$.
The panels on the left side show optimal segment duration
$\Tseg$ and number of segments $\Nseg$, and the panels on the right
side show optimal coherent $\co{m}$ and semicoherent $\ic{m}$
mismatch.  Panels (a) and (b) are obtained using greedy data selection,
the most sensitive optimal solution is
 \input{DC70DS1_optimal.tex}. Panels (c) and (d) are
 obtained using compact data selection, the most sensitive optimal
 solution is \input{DC70DS0_optimal.tex}. Panels (e) and (f)
are obtained using greedy-compact data selection, the most sensitive
optimal solution is \input{DC70DS2_optimal.tex}}
\label{fig:DC70}
\end{figure*}

We consider now data with gaps allowing a duty cycle (fraction
 of actually available data $\Tdata$ in a given time span $\Tspan$, $\epsilon\equiv\Tdata/\Tspan$) of $70\ \%$ per detector,
 while the noise floor is still constant $\sqrt{\Sn} = 1\ \Hz^{-1/2}$. 
The results of the numerical optimization are plotted in
 Fig.~\ref{fig:DC70}, whereas the optimal solutions are summarized
in Table \ref{tab:caseB}. 

\begin{table}[h!]
\begin{tabular}{|c|c|c|c|}\hline
 & greedy & compact & greedy-compact \\\hline\hline
$\sensdep\,[\Hz^{-1/2}]$ & $64.8$ & $68.0$ & $65.0$\\
$T\,[\days]$ & 364.3& 220.5 & 364.7 \\
$\Nseg$ & 326 & 165.8 & 434.1 \\
$\Tseg\,[\days]$ & 0.9 & 1.3 & 0.7 \\
$\co{m}$ & 0.07 & 0.08 & 0.04 \\
$\ic{m}$ & 0.46 & 0.27 & 0.35 \\\hline
\end{tabular}
\caption{Optimal solution using greedy, compact and
 greedy-compact data selection applied to data from 2 detectors,
with 70 \% duty cycle and of constant noise floor $\sqrt{\Sn}=1\ \Hz^{-1/2}$.}
 \label{tab:caseB}
 \end{table}

\subsubsection*{Data with gaps  and noise floor fluctuations}
\begin{figure*}[htbp]
  \subfigure[ ]{\includegraphics[width=0.49\linewidth]
{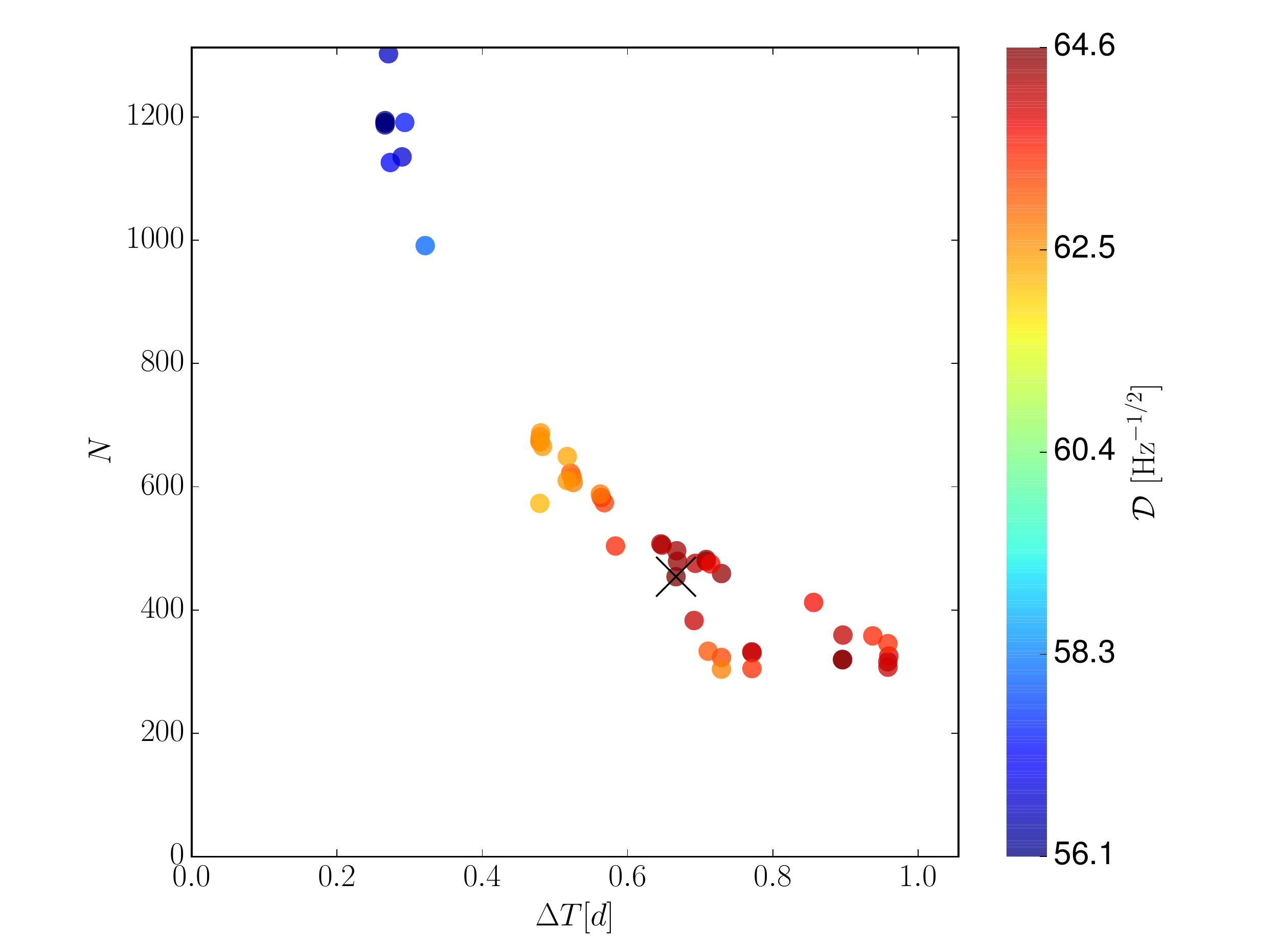}
\label{fig:DC70NDS1_hnot}}
 \subfigure[ ]{\includegraphics[width=0.49\linewidth]{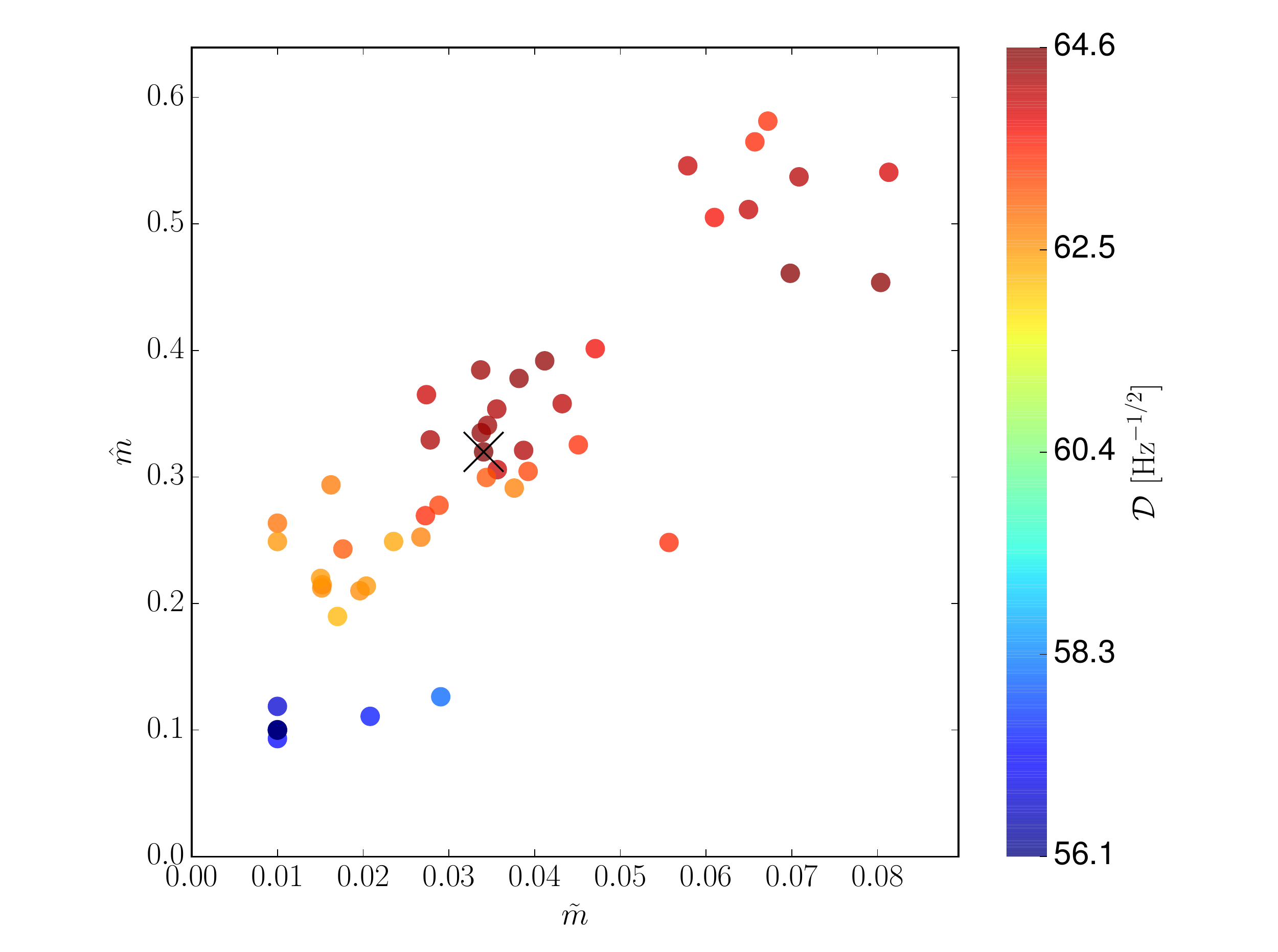}\label{
fig:DC70NDS1_m}}
\quad
  \subfigure[ ]{\includegraphics[width=0.49\linewidth]
{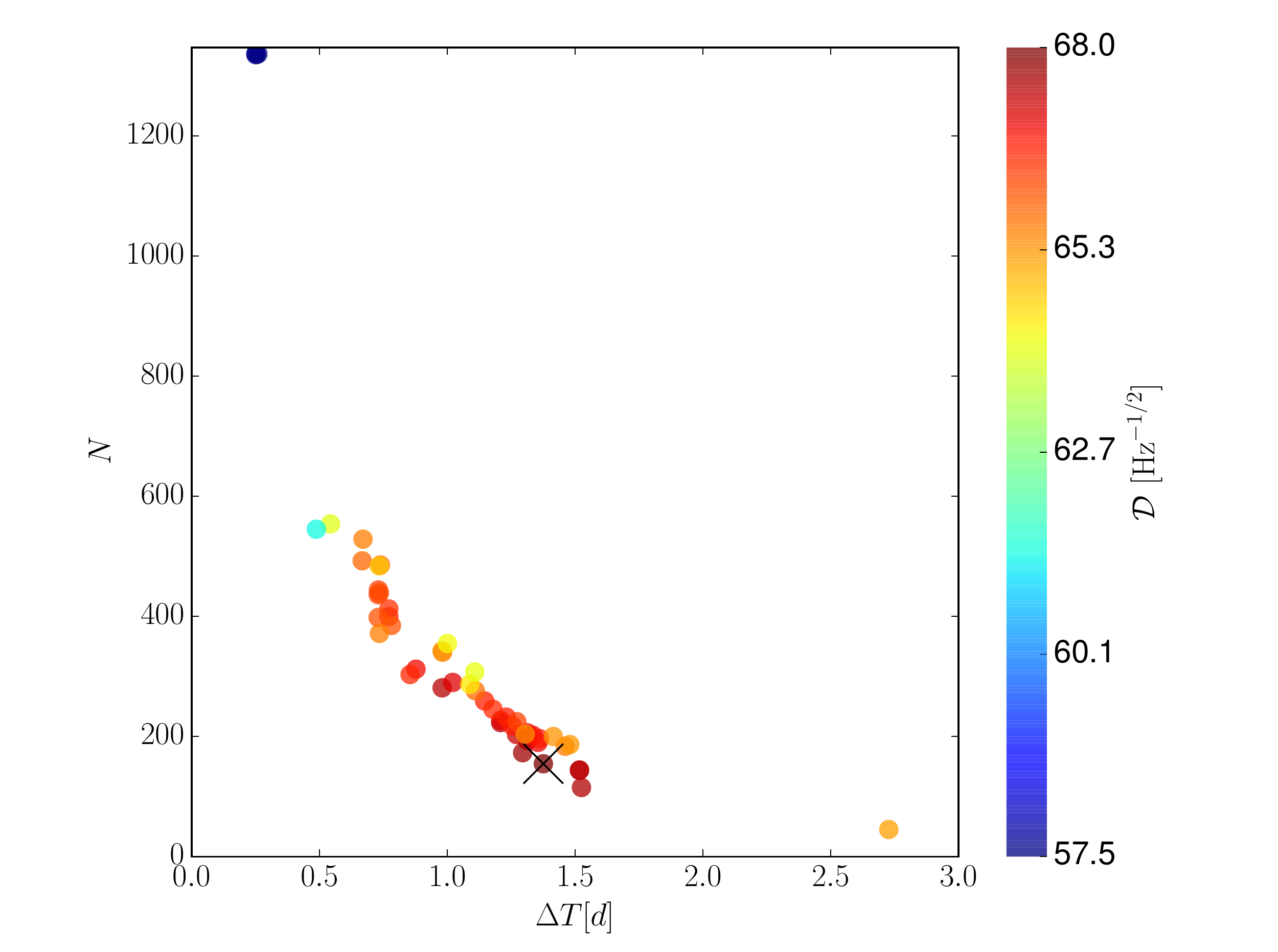}
\label{fig:DC70NDS0_hnot}}
 \subfigure[ ]{\includegraphics[width=0.49\linewidth]{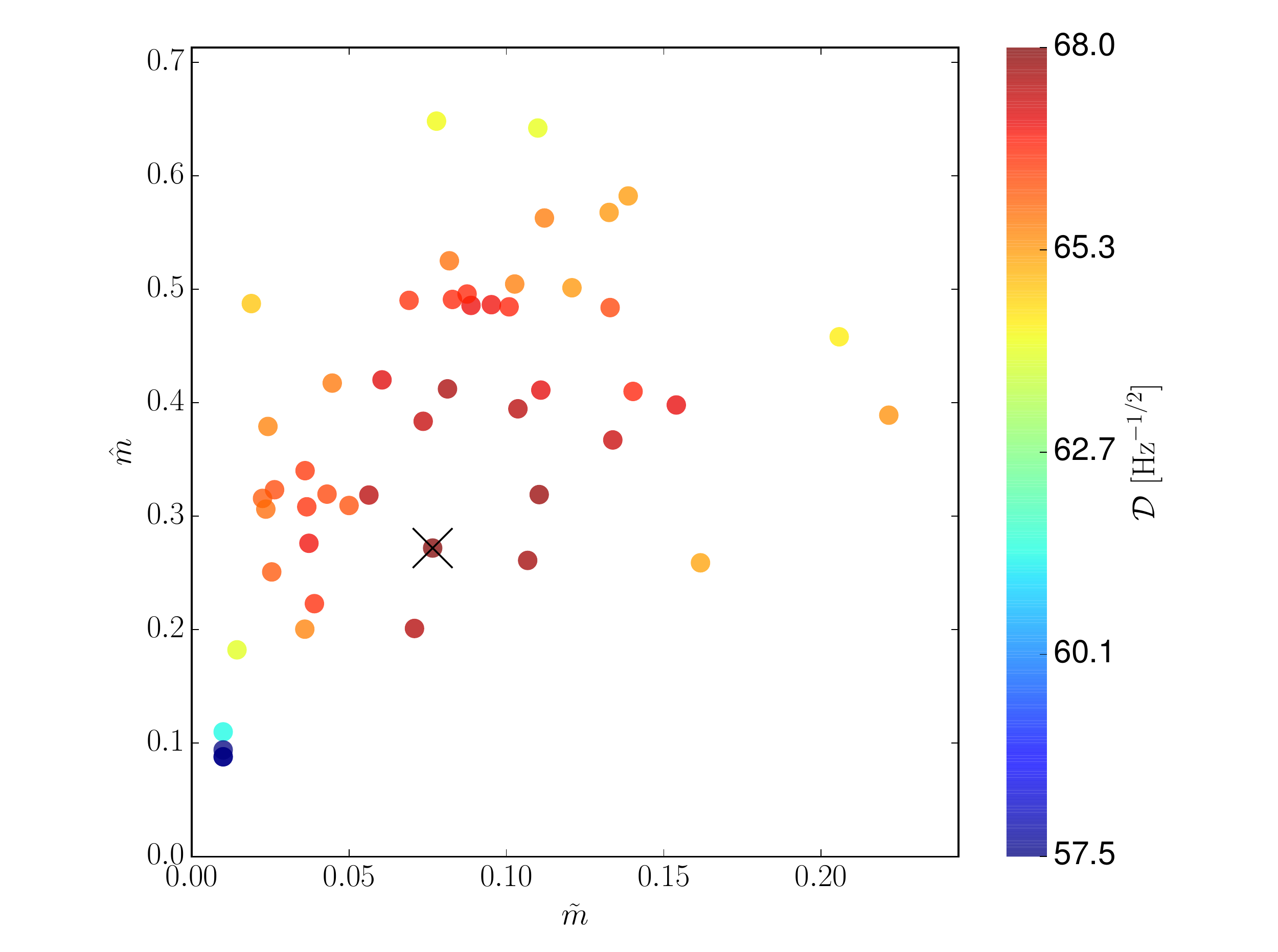}\label{
fig:DC70NDS0_m}}
\quad
  \subfigure[ ]{\includegraphics[width=0.49\linewidth]
{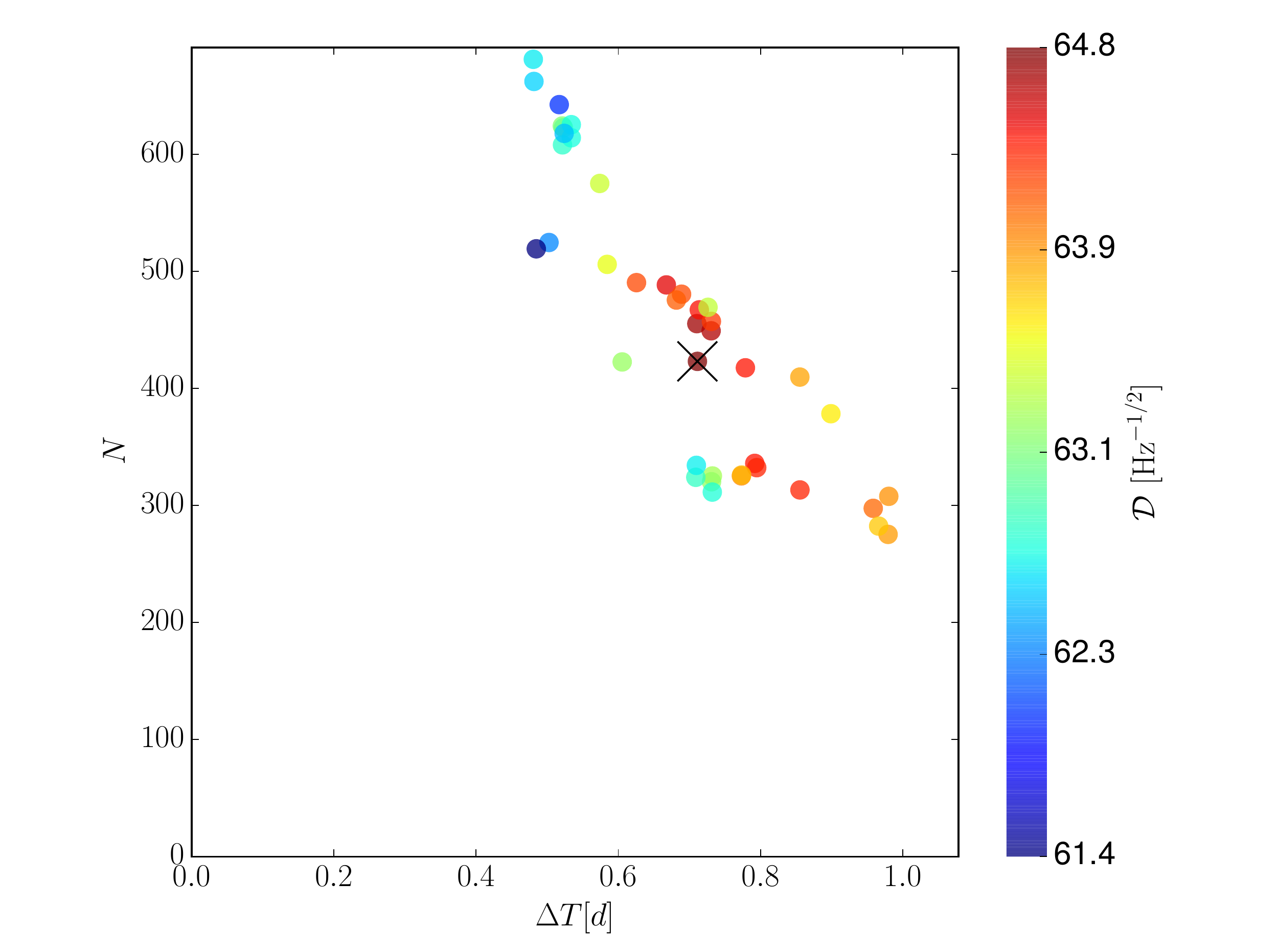}
\label{fig:DC70NDS2_hnot}}
 \subfigure[ ]{\includegraphics[width=0.49\linewidth]{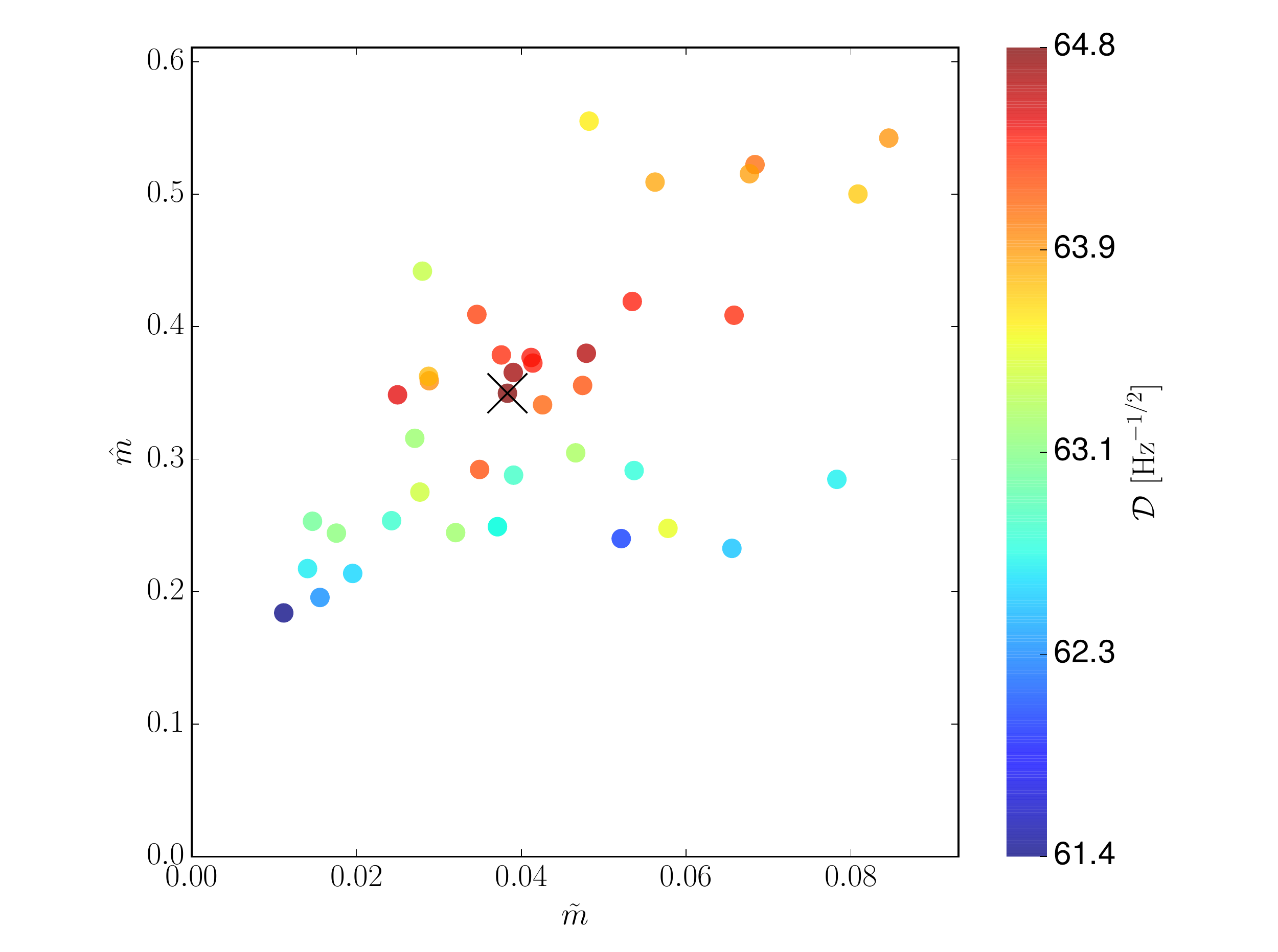}\label{
fig:DC70NDS2_m}}
\caption[Numerical optimization using simulated data with gaps.]{Semicoherent search
 optimization for data from 2 detectors with duty
 cycle 70\% and noise floor with fluctuations. The cost constraint is
 $\CC_{0}=472.0\,\days$. The best  numerical solution is denoted with
 $\times$.
The panels on the left side show optimal segment duration
$\Tseg$ and number of segments $\Nseg$, and the panels on the right
side show optimal coherent $\co{m}$ and semicoherent $\ic{m}$
mismatch.  Panels (a) and (b) are obtained using greedy data selection,
the most sensitive optimal solution is
 \input{DC70NDS1_optimal.tex}. Panels (c) and (d) are
 obtained using compact data selection, the most sensitive optimal
 solution is \input{DC70NDS0_optimal.tex}. Panels (e) and (f)
are obtained using greedy-compact data selection, the most sensitive
optimal solution is \input{DC70NDS2_optimal.tex}}
\label{fig:DC70N}
\end{figure*}

In this example we further relax the requirements on the data by
 allowing noise floor fluctuations, while keeping the duty cycle of  
$70\ \%$ per detector.  For each SFT  the PSD has been drawn from a
 Gaussian distribution with mean $E[\sqrt{\Sn}]=1\ \Hz^{-1/2}$ and
standard deviation $\sigma[\sqrt{\Sn}]=15\times10^{-2}\ \Hz^{-1/2}$.
The outcome of the optimization  is plotted in  Fig.~\ref{fig:DC70N}. 
The optimal parameters are summarized in Table \ref{tab:caseC}.
 
\begin{table}[htb]
\begin{tabular}{|c|c|c|c|}\hline
 & greedy & compact & greedy-compact \\\hline\hline
$\sensdep\,[\Hz^{-1/2}]$ & $64.8$ & $68.0$ & $64.8$\\
$T\,[\days]$ & 364.9 & 215.3 & 364.8 \\
$\Nseg$ & 454.1 & 154.6  & 423 \\
$\Tseg\,[\days]$ & 0.7  & 1.4 & 0.7  \\
$\co{m}$ & 0.03 & 0.08 & 0.04  \\
$\ic{m}$ & 0.32  & 0.27  & 0.35  \\\hline
\end{tabular}
\caption{Optimal solution using greedy, compact and
 greedy-compact data selection applied to data from 2 detectors,
with 70 \% duty cycle and noise floor with fluctuations.}
 \label{tab:caseC}
 \end{table}

\subsection{Directed search using real data}

In this subsection we apply the optimization procedure to real data
 collected by the Hanford (H1) and Livingston (L1) LIGO detectors
 during the S5 run \cite{Abbott:2007kv}.  The most sensitive data is found around 
$169.875\ \Hz$, thus the optimization will be done at this frequency. 
The details about the data are summarized in Table \ref{tab:data}. 
It spans 653 days in  17797 SFTs of duration $\Tsft = 1800\ \secs$. 
With this  the average duty cycle is approximately $0.28$ in each
 detector.

\begin{table}[htbp]
 \centering
\begin{tabular}[c]{|c|c|c|c|c|c|c|}\hline
run & detector & $f$ [Hz] & first SFT & last SFT & $\Nsft$ & $\Tspan$ [d]
\\\hline\hline
S5 & H1 & 169.875 & 818845553 & 875277921 & 9331 & 653 \\\hline
S5 & L1 & 169.875 & 818845553 & 875278812 & 8466 & 653\\\hline
\end{tabular}
\caption{Detector data used to test the numerical optimization under real conditions.}
\label{tab:data}
\end{table}

\subsubsection*{Keeping the cost constraint}
\begin{figure*}[htbp]
\centering
  \subfigure[ ]{\includegraphics[width=0.49\linewidth]{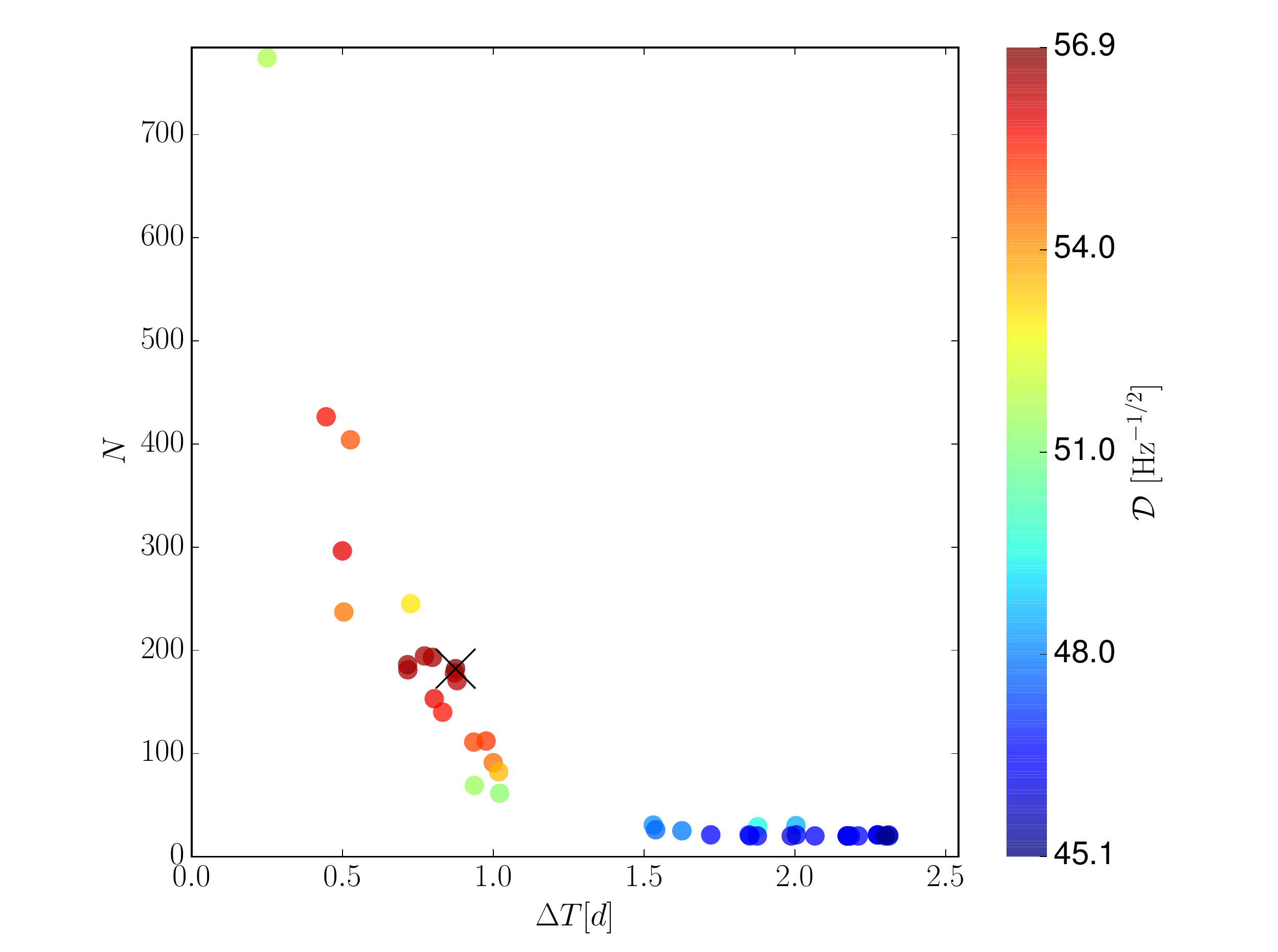}\label{fig:S5DS1_hnot}}
  \subfigure[ ]{\includegraphics[width=0.49\linewidth]{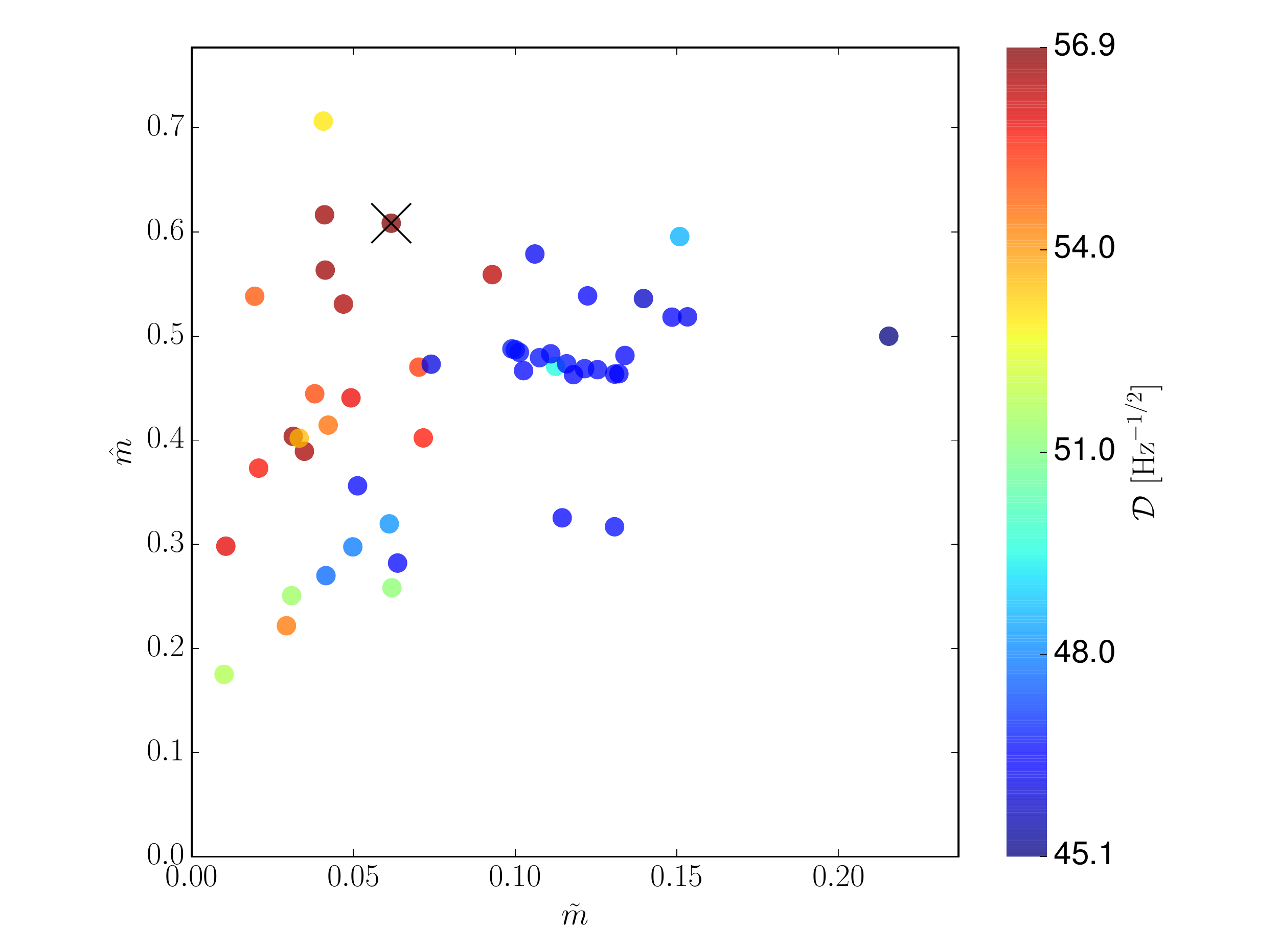}\label{fig:S5DS1_m}}
\quad
  \subfigure[ ]{\includegraphics[width=0.49\linewidth]{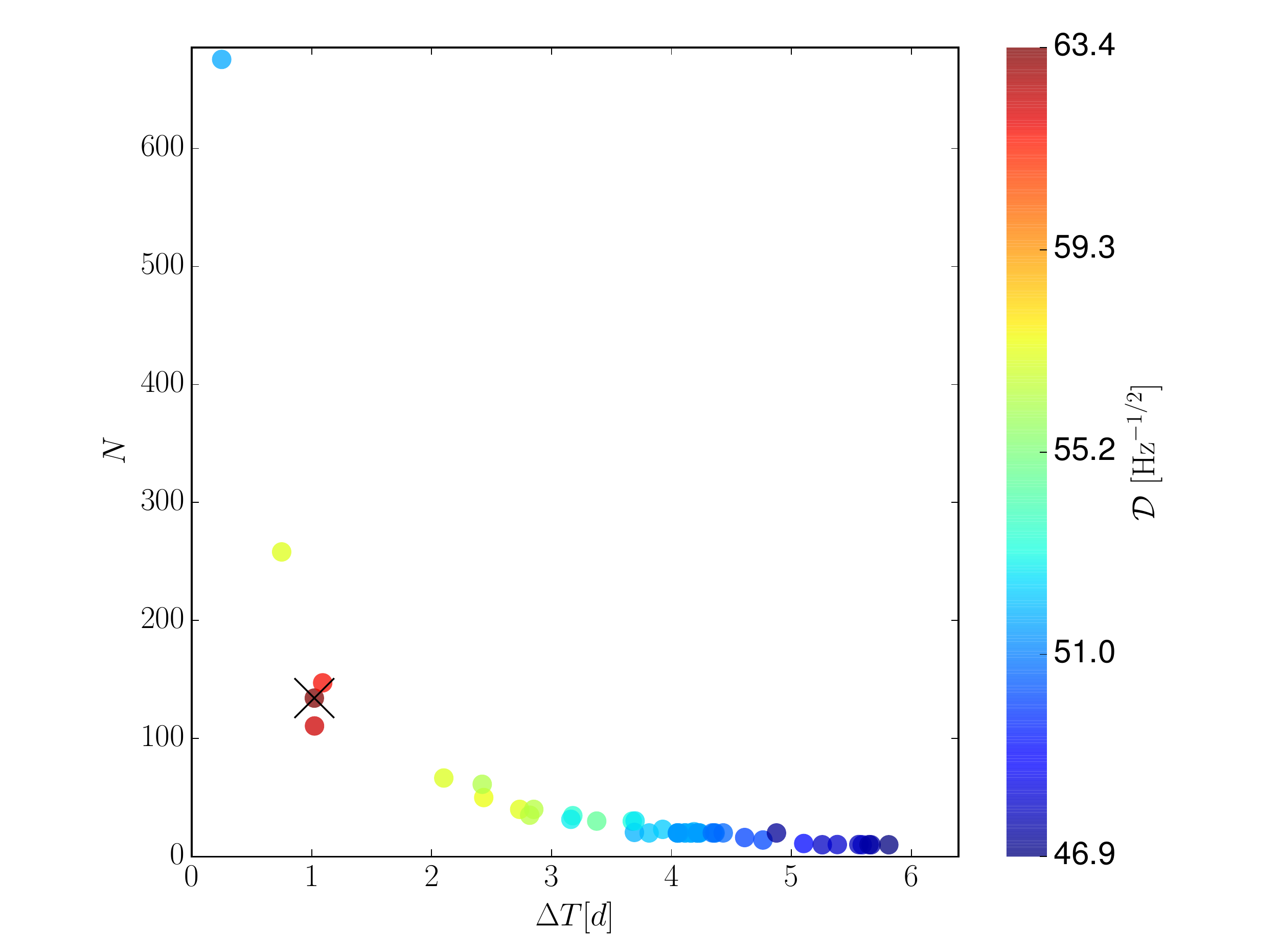}\label{fig:S5DS0_hnot}}
  \subfigure[ ]{\includegraphics[width=0.49\linewidth]{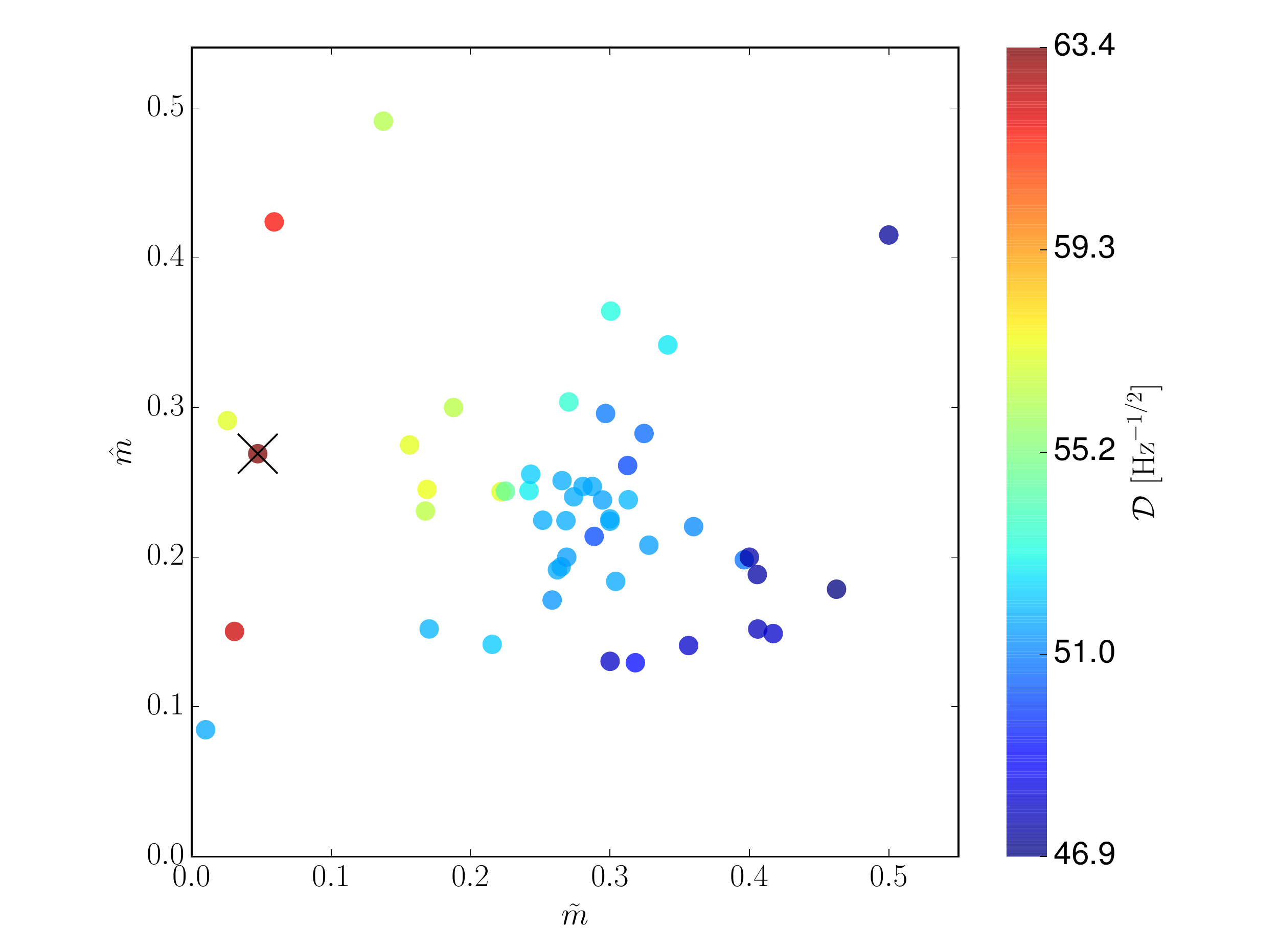}\label{fig:S5DS0_m}}
\quad
  \subfigure[ ]{\includegraphics[width=0.49\linewidth]{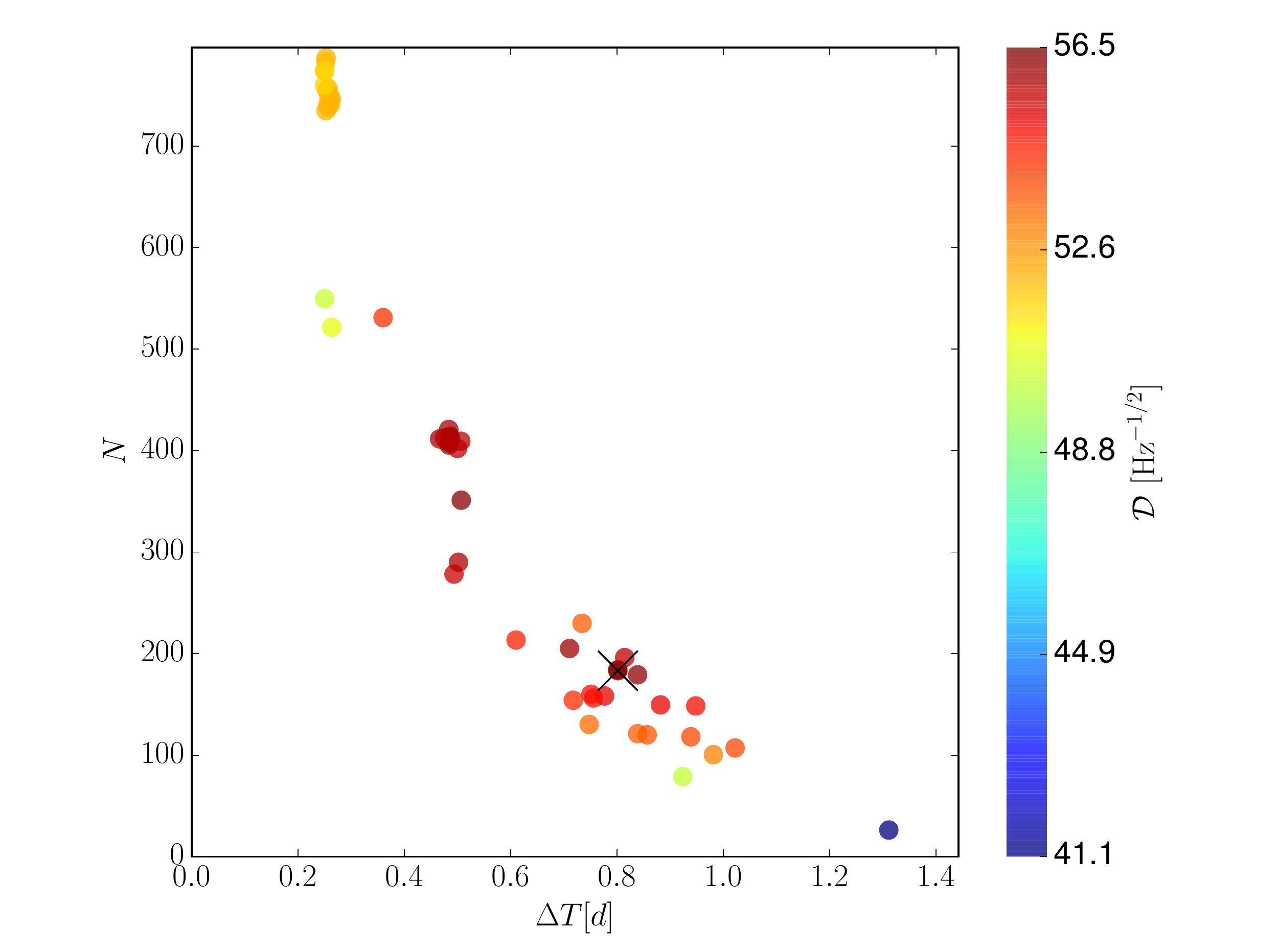}\label{fig:S5DS2_hnot}}
  \subfigure[ ]{\includegraphics[width=0.49\linewidth]{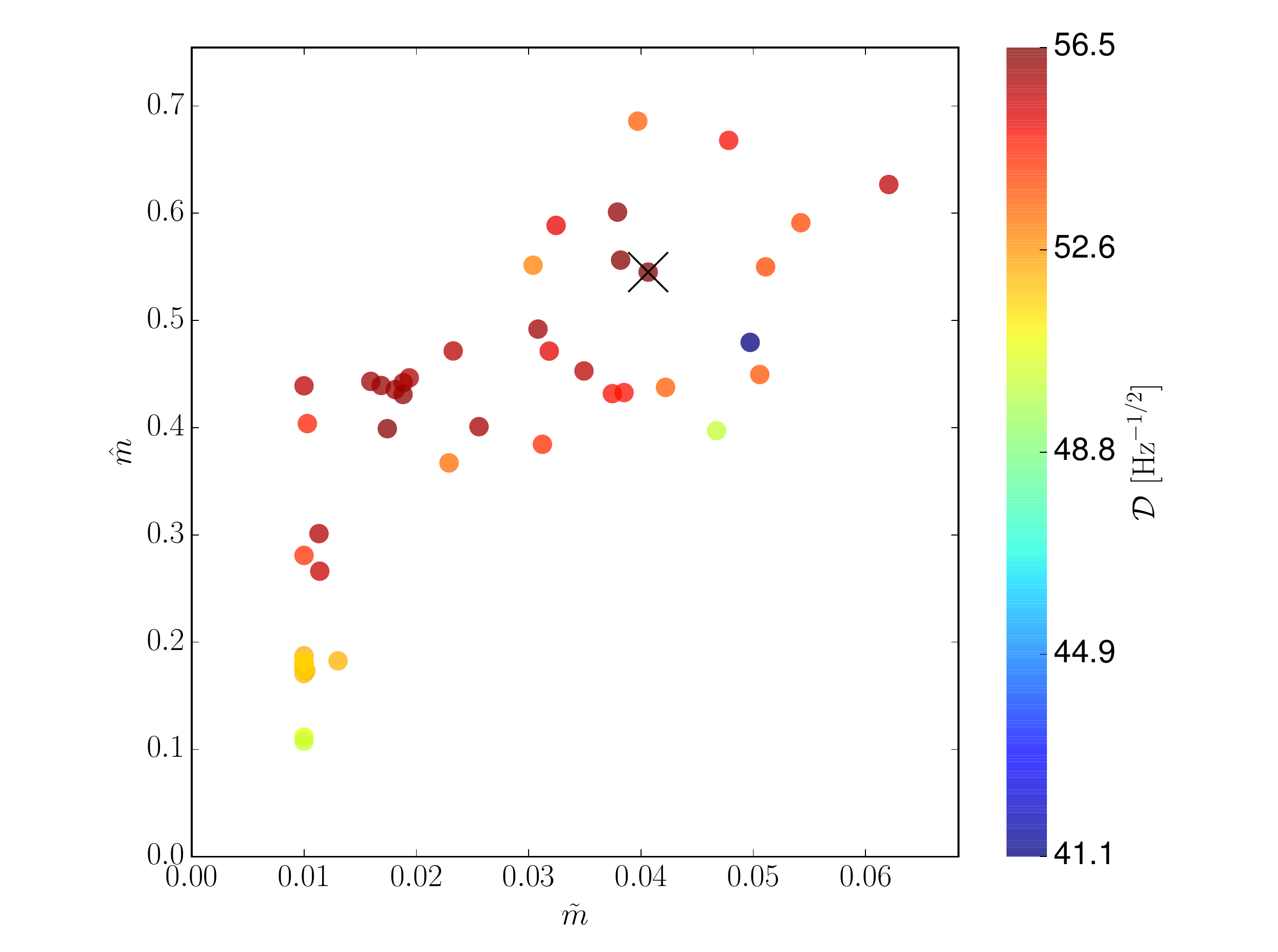}\label{fig:S5DS2_m}}
\caption[Numerical optimization using real data.]{Semicoherent search optimization for data from the H1 and L1 LIGO detectors
during the S5 run at $169.875\ \Hz$. The cost constraint is $\CC_{0}=472.0\
\days$.  The best numerical solution is denoted with $\times$. 
The panels on the left side show optimal segment duration
$\Tseg$ and number of segments $\Nseg$, and the panels on the right
side show optimal coherent $\co{m}$ and semicoherent $\ic{m}$
mismatch.  Panels (a) and (b) are obtained using greedy data selection,
the most sensitive optimal solution is
 \input{S5DS1_optimal.tex}. Panels (c) and (d) are
 obtained using compact data selection, the most sensitive optimal
 solution is \input{S5DS0_optimal.tex}. Panels (e) and (f)
are obtained using greedy-compact data selection, the most sensitive
optimal solution is \input{S5DS2_optimal.tex}}
\label{fig:S5}
\end{figure*}

We first keep the cost constraint equal to the computing cost used  in
 the examples with simulated data, namely $\CC_{0}\approx472$ days. 
The result of the optimization procedure is plotted in Fig.~\ref{fig:S5}. 
We summarize the optimal parameters in Table \ref{tab:caseD}. In this
case usage of the compact data selection algorithm yields approximately 
10\% increase of the search depth compared to the other two methods. 
The gain of search depth compared to the fully-coherent solution is 
approximately $1.5\ $.

\begin{table}[htb]
\begin{tabular}{|c|c|c|c|}\hline
 & greedy & compact & greedy-compact \\\hline\hline
$\sensdep\,[\Hz^{-1/2}]$ & $56.9$ & $63.4$ & $56.5$\\
$T\,[\days]$ & 554.4 & 329.0 & 653.0 \\
$\Nseg$ & 182.2 & 134.3  & 183.1 \\
$\Tseg\,[\days]$ & 0.9 & 1.0 & 0.8  \\
$\co{m}$ & 0.06 & 0.05 & 0.04 \\
$\ic{m}$ & 0.61 & 0.27 & 0.55  \\\hline
\end{tabular}
\caption{Optimal solution using greedy, compact and
 greedy-compact data selection applied to data from the H1 and L1
LIGO detectors during the S5 run.}
 \label{tab:caseD}
 \end{table}


\subsubsection*{Using Einstein@Home}

Finally we consider using the Einstein@Home distributed computing
 environment to increase the computing cost constraint to
 $\CC_{0}=360\times10^{3}$ days
on a single computing core.  Such computing power corresponds to
 approximately 30 days on 12000 24x7 single core CPUs. 

\begin{table}[htb]
\begin{tabular}{|c|c|c|c|}\hline
 & greedy & compact & greedy-compact \\\hline\hline
$\sensdep\,[\Hz^{-1/2}]$ & $98.6$ & $97.9$ & $95.6$\\
$T\,[\days]$ & 464.8 & 387.7 & 653.2 \\
$\Nseg$ & 20 & 20  & 35.2 \\
$\Tseg\,[\days]$ & 15.4 & 17.9 &  9.8 \\
$\co{m}$ & 0.10 & 0.11 & 0.04 \\
$\ic{m}$ & 0.28 & 0.29 & 0.26  \\\hline
\end{tabular}
\caption{Optimal solution using greedy, compact and
 greedy-compact data selection applied to data from the H1 and L1
LIGO detectors during the S5 run. The cost constraint is suitable
with Einstein@Home.}
 \label{tab:caseE}
 \end{table}

\begin{figure*}[htbp]
\centering
  \subfigure[ ]{\includegraphics[width=0.49\linewidth]{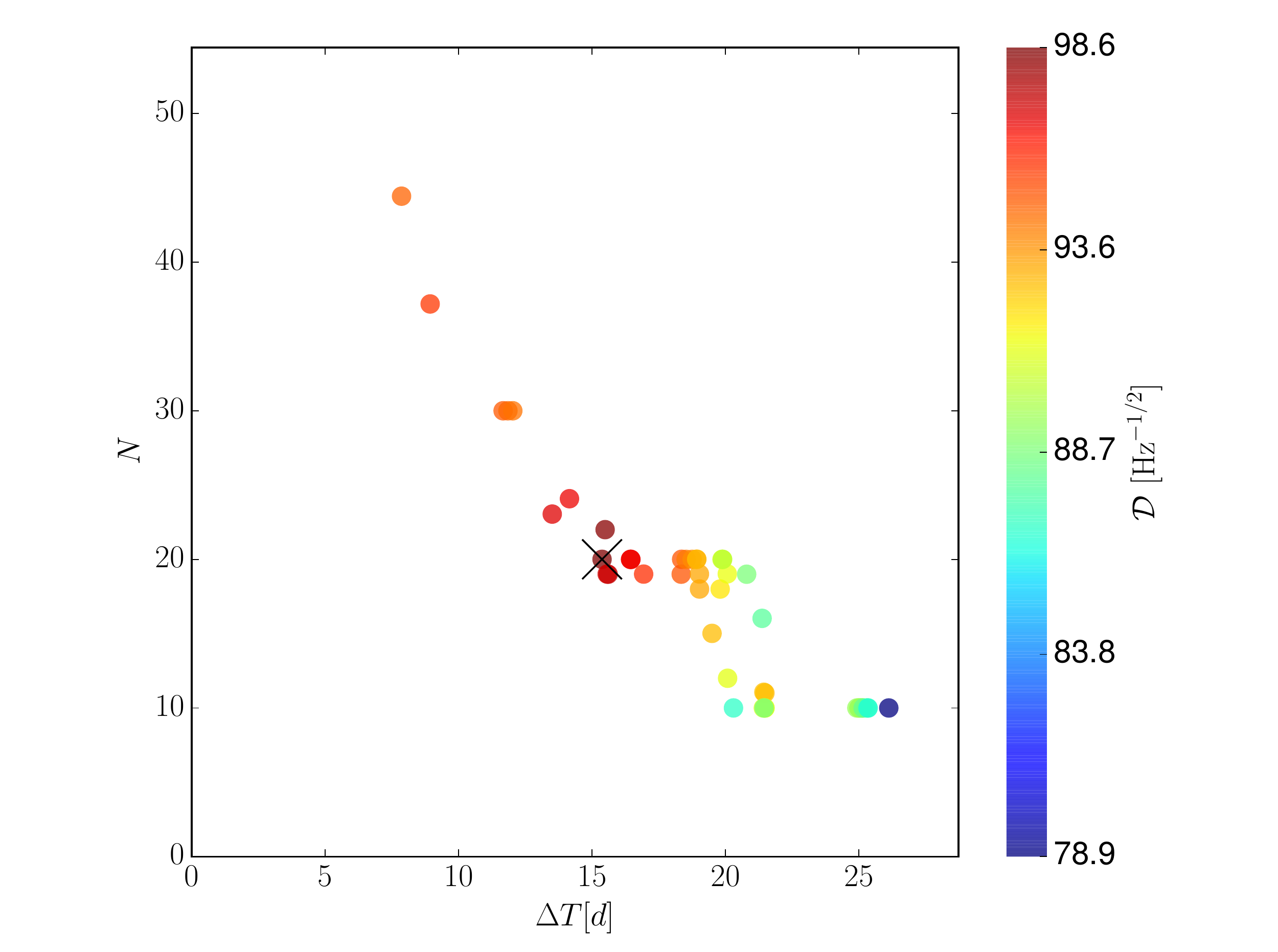}
\label{fig:S5EAHDS1_hnot}}
  \subfigure[ ]{\includegraphics[width=0.49\linewidth]{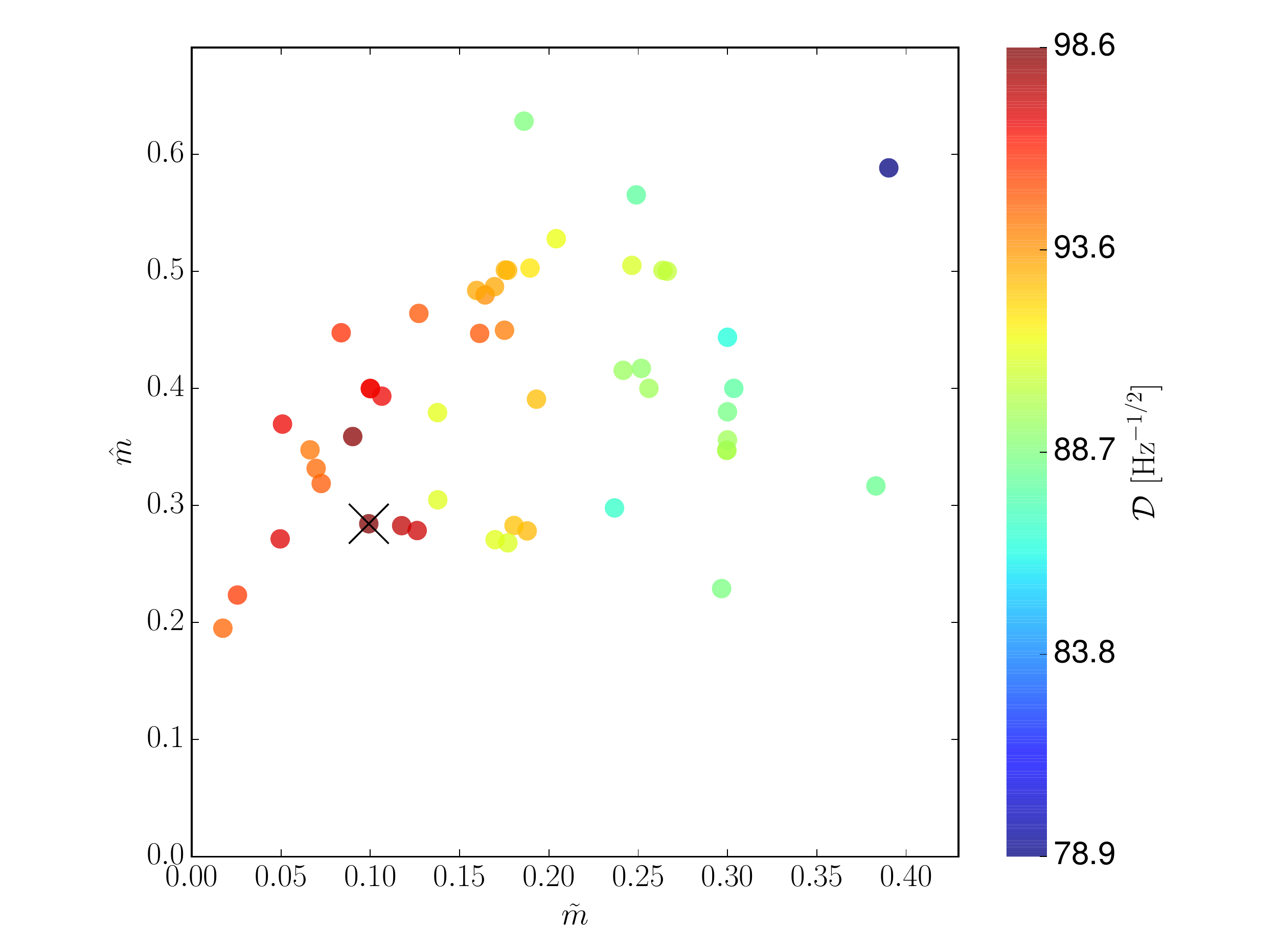}\label{
fig:S5EAHDS1_m}}
\quad
  \subfigure[ ]{\includegraphics[width=0.49\linewidth]{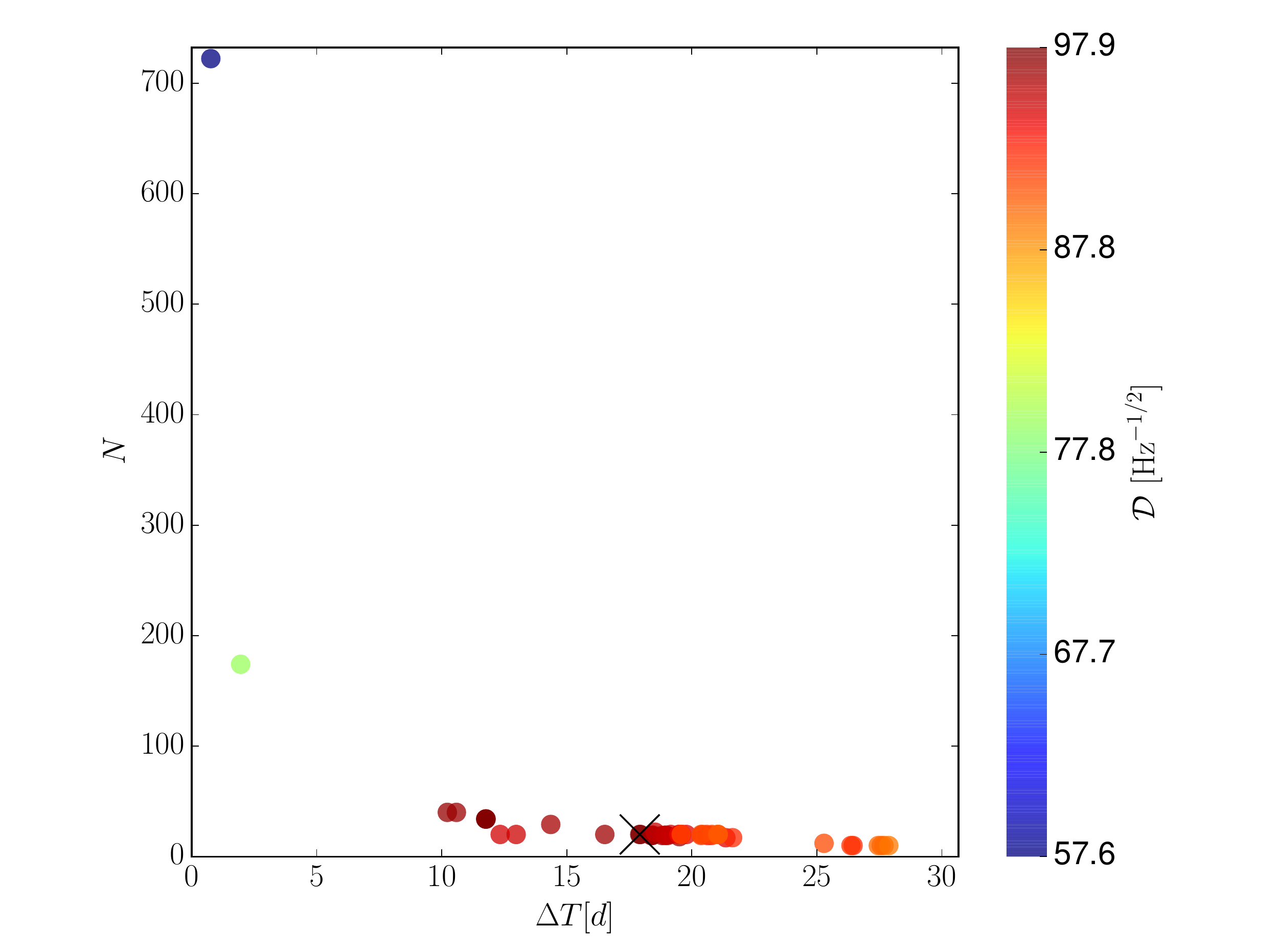}
\label{fig:S5EAHDS0_hnot}}
  \subfigure[ ]{\includegraphics[width=0.49\linewidth]{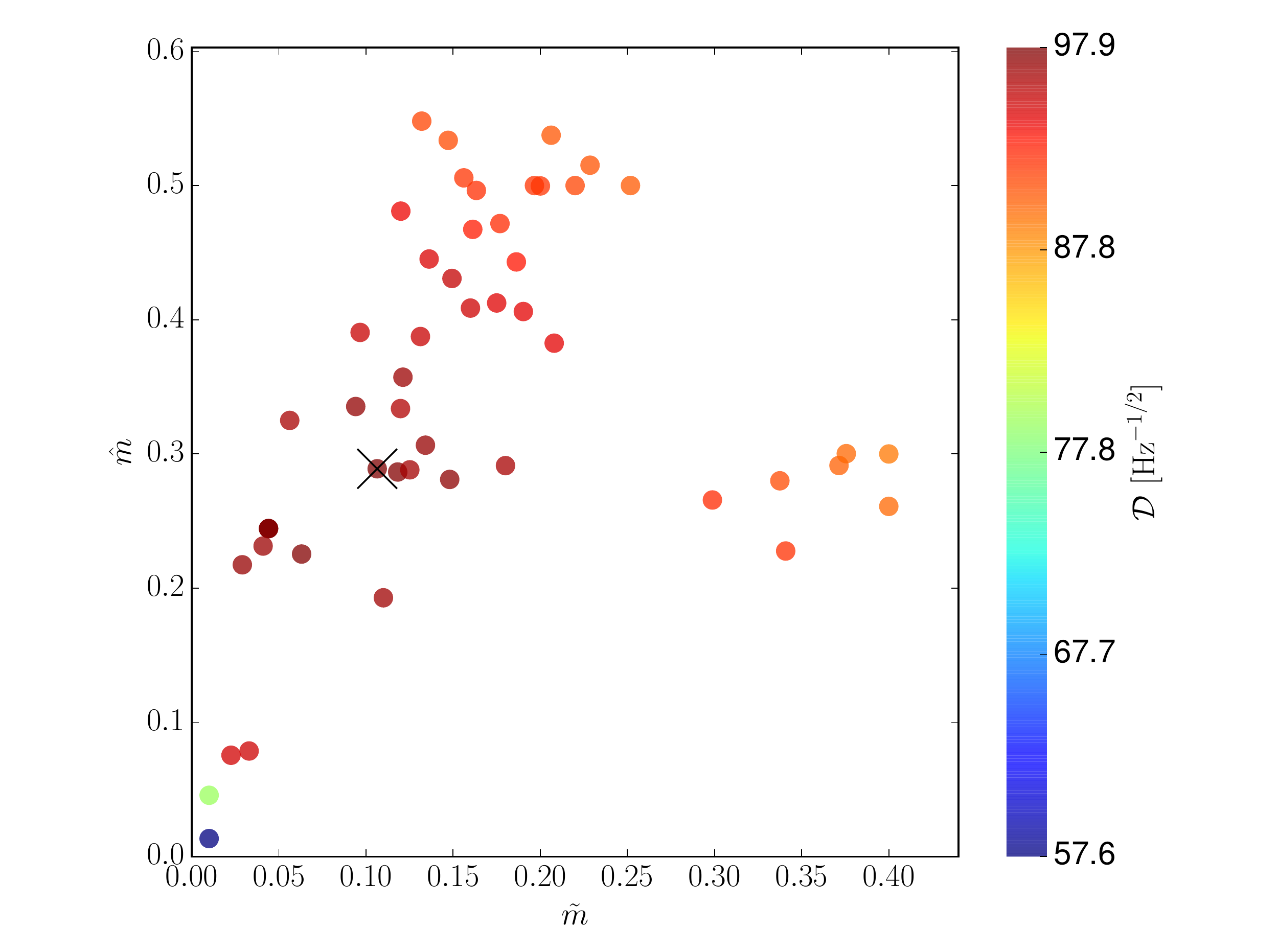}\label{
fig:S5EAHDS0_m}}
\quad
  \subfigure[ ]{\includegraphics[width=0.49\linewidth]{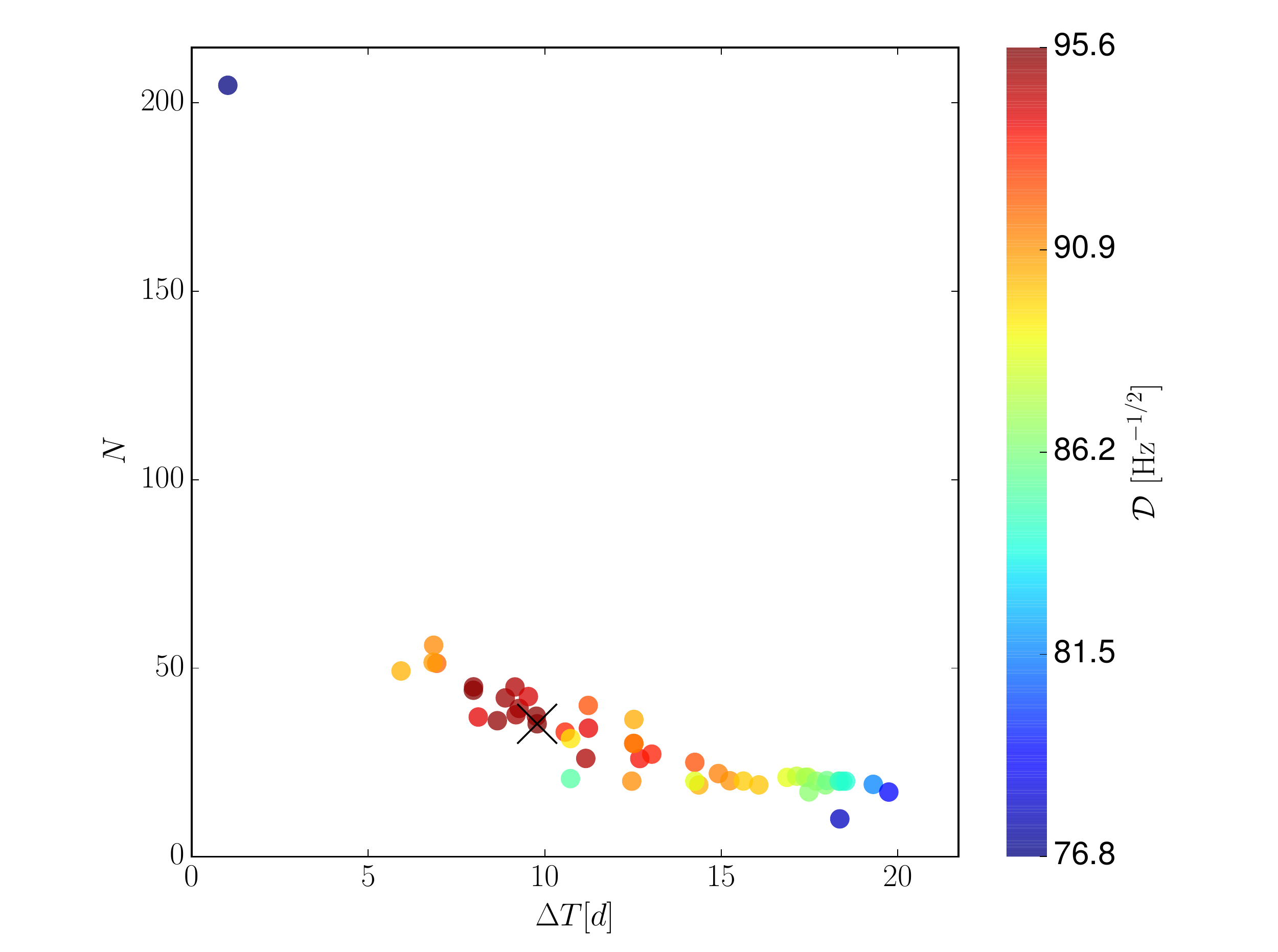}
\label{fig:S5EAHDS2_hnot}}
  \subfigure[ ]{\includegraphics[width=0.49\linewidth]{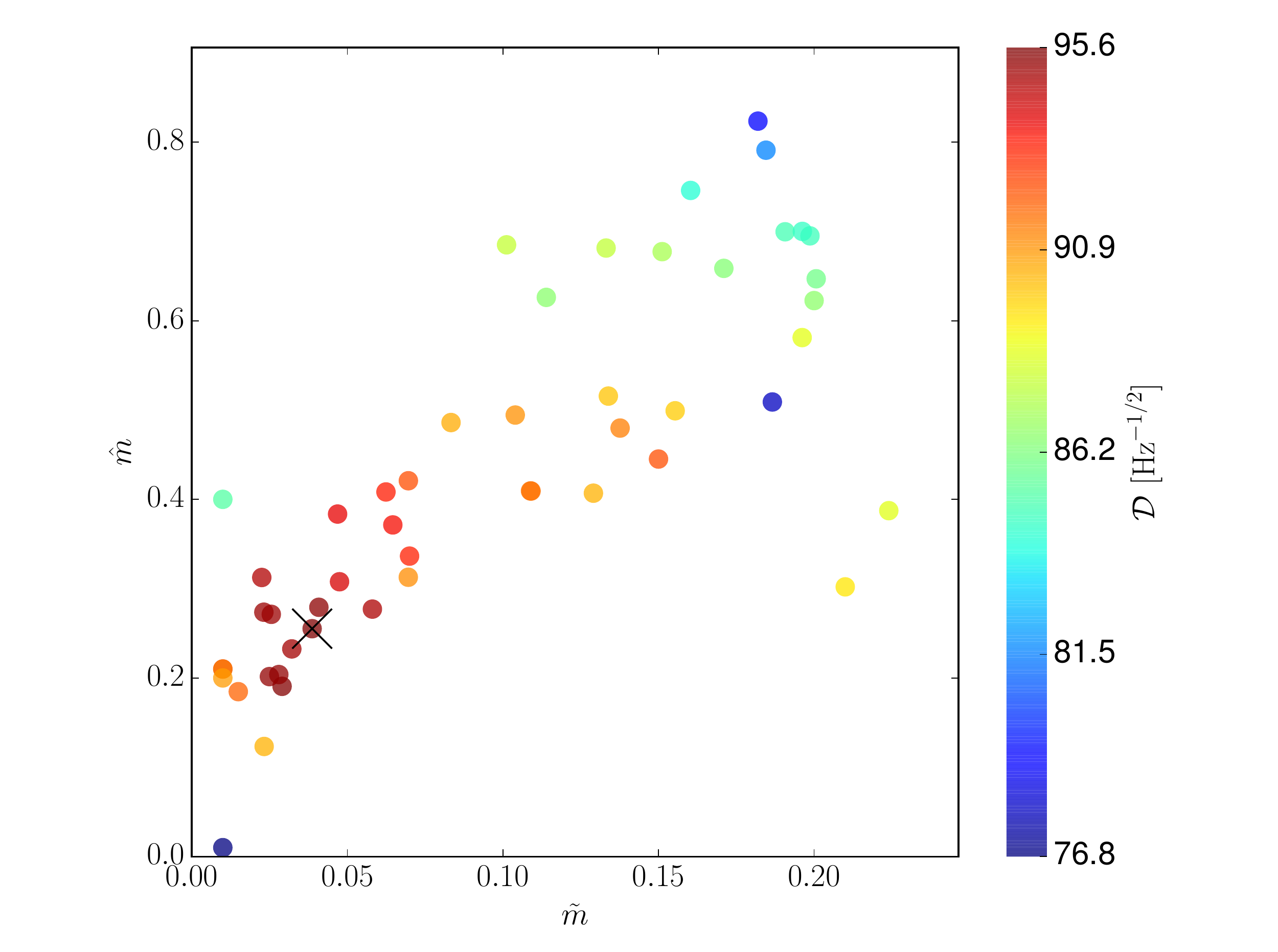}\label{
fig:S5EAHDS2_m}}
\caption[Numerical optimization using real data for an Einstein@Home search.]
{Semicoherent search optimization with data from
 the H1 and L1 LIGO detectors during
the S5 run at $169.875\ \Hz$. The cost constraint is 
$\CC_{0}=360\times10^{3}\ \days$. The best numerical solution is denoted
 with $\times$. 
The panels on the left side show optimal segment duration
$\Tseg$ and number of segments $\Nseg$, and the panels on the right
side show optimal coherent $\co{m}$ and semicoherent $\ic{m}$
mismatch.  Panels (a) and (b) are obtained using greedy data selection,
the most sensitive optimal solution is
 \input{S5EAHDS1_optimal.tex}. Panels (c) and (d) are
 obtained using compact data selection, the most sensitive optimal
 solution is \input{S5EAHDS0_optimal.tex}. Panels (e) and (f)
are obtained using greedy-compact data selection, the most sensitive
optimal solution is \input{S5EAHDS2_optimal.tex}}
\label{fig:S5EAH}
\end{figure*}

The results of the numerical optimization procedure  approximation are
 plotted in Fig.~\ref{fig:S5EAH}. The optimal parameters are given in
 Table \ref{tab:caseE}.

Note, that by the enormous increase of the
 computing power, the three different data selection methods yield practically 
 equal sensitivity and the gain in sensitivity compared to the
 fully coherent solution is approximately $2.65\ $.

\section{Discussion}
\label{sec:5}
 In this paper we studied the optimization of semicoherent searches for continuous
gravitational waves, in particular the StackSlide search, at constrained 
computing cost under more realistic conditions by taking into account 
possible gaps in the data and noise level changes. The presented method to obtain optimal
search parameters is based on numerical optimization combined with a data 
selection algorithm. The outcome of the optimization procedure is the
 set of the optimal search parameters $\{\co{m}\ ,\ic{m}, \Nseg,\ \Tseg\}$
as well as the selected data and an estimate of the expected sensitivity depth.

We showed that under ideal conditions, our numerical optimization method 
recovers, in terms of sensitivity, the optimal solution found by using the analytical method 
discussed in Ref. \cite{PrixShaltev2011}. Based on the examples of practical 
application, we conclude that the compact data selection yields higher sensitivity 
depth compared to the greedy data selection. However the superiority of the 
compact method over the greedy method depends on the data quality and 
on the computing cost constraint. It diminishes namely for data without large 
differences in the noise level from epoch to epoch and equally distributed 
gaps or for large allowed computing cost, where we can use nearly all the 
data.
 
The optimization procedure is immediately applicable 
to searches over simple (nearly) ''box``  parameter-space shape.
While the proposed optimization method can be easily adapted to
 other  types of searches, by modification of the computing cost
 function, further work is required to extend the applicability 
of the optimization procedure to an arbitrary parameter-space shape.
The proposed optimization method assumed a fixed frequency band.
Further work is required to relax this condition, in order to
answer the question, what is the optimal trade-off between
the size of the searched parameter space (width of the search)
and its coverage (depth of the search). For a promising approach 
see \cite{Ming:2015jla}. In the example section of this paper we 
considered directed searches. For further work on all-sky searches 
one should take into account recent research on the semicoherent metric 
\cite{PhysRevD.92.082003}, as it suggests increase of the semicoherent 
number of templates with yet unknown implications.

\section{Acknowledgments}
MS would like to acknowledge the numerous comments and suggestions
of Reinhard Prix. The author is also thankful for the discussions with Karl Wette, 
Badri Krishnan, Sinead Walsh and Maria Alessandra Papa. MS gratefully 
acknowledges the support of Bruce Allen and the IMPRS on Gravitational Wave 
Astronomy of the Max-Planck-Society. The numerical optimizations in this
work have been performed on the ATLAS computing cluster of the 
Max-Planck-Institut f\"ur Gravitationsphysik. This paper has been assigned LIGO document number \dcc.

\appendix
\section{Pseudo code of the proposed data selection algorithms}
In the following we describe the data selection algorithms proposed in 
Sec. \ref{sec:3} in pseudo code.

\begin{algorithm}[H]
\caption {Greedy data selection.}
\begin{algorithmic}[0]
\Input{$\sftlist$ - list of SFTs sorted by increasing timestamp $t_{j}$, 
$\Nseg$ - requested number of segments with duration $\Tseg$}\EndInput
\Output{$\seglist$ - list of segments and corresponding SFTs}\EndOutput

\While{\Call{Length}{$\seglist$}$<\Nseg$ \&\& \Call{Length}{$\sftlist$} $> 0$}
\State $\gudlist\gets0$\Comment{List of goodness per segment}
\ForAll{$t_{j}$}
 \State $d\gets$\Call{FindAllSFTsInRange}{$t_{j},t_{j}+ \Tseg$}
 \State $\mG_{j}\gets$\Call{ComputeGoodness}{d}
\EndFor
\State $g\gets$\Call{Max}{$\gudlist$}\footnote{Working with real numbers it is unlikely to find segments with exactly
 equal goodness, but in such rare cases we choose the earliest segment in time.}
\State $\seglist_{j}\gets$\Call{AddSegment}{g}
\State \Call{RemoveUsedSFTs}{$\seglist_{j}$}
\EndWhile
\end{algorithmic}
\label{alg:greedy-data-selection}
\end{algorithm}

\begin{algorithm}[H]
\caption {Compact data selection.}
\begin{algorithmic}[0]
\Input{$\sftlist$ - list of SFTs sorted by increasing timestamp $t_{j}$, 
$\Nseg$ - requested number of segments with duration $\Tseg$}\EndInput
\Output{$\seglist$ - list of segments and corresponding SFTs}\EndOutput
\State $\gudlist\gets0$\Comment{List of goodness}
\State $L\gets0$\Comment{List of segmentlists}
\ForAll{$t_{j}$}
\State $t_{s}\gets t_{j}$
\State $l\gets0$
\While{\Call{Length}{$l$}$<\Nseg$ \&\& $t_{s}<$\Call{Max}{$t_{j}$}}
 \State $d\gets$\Call{FindAllSFTsInRange}{$t_{s},t_{s}+ \Tseg$}
 \State $l_{j}\gets$\Call{AddSegment}{d}
 \State $t_{s}\gets$\Call{FirstTimestampAfter}{$t_{s}+\Tseg$}
\EndWhile
 \State $L_{j}\gets l$
 \State $\mG_{j}\gets$\Call{ComputeGoodness}{$L_{j}$}
\EndFor
\State $\seglist\gets$\Call{Max}{$\gudlist$}\Comment{such that the computing cost 
constraint is satisfied}
\end{algorithmic}
\label{alg:compact-data-selection}
\end{algorithm}

\begin{algorithm}
\caption {Greedy-compact data selection.}
\begin{algorithmic}[0]
\Input{$\sftlist$ - list of SFTs sorted by increasing timestamp $t_{j}$,
$\Nseg$ - requested number of segments with duration $\Tseg$}\EndInput
\Output{$\seglist$ - list of segments and corresponding SFTs}\EndOutput

\While{\Call{Length}{$\seglist$}$<\Nseg$ \&\& \Call{Length}{$\sftlist$} $> 0$}
\State $\gudcost\gets0$\Comment{List of goodness / cost ratio}
\ForAll{$t_{j}$}
 \State $d\gets$\Call{FindAllSFTsInRange}{$t_{j},t_{j}+ \Tseg$}
 \State $\gudcost_{j}\gets$\Call{ComputeGoodnessCostRatio}{d}\footnote{We compute 
 $\mG_{j}/\CC_{j}^{2}$, where $\CC_{j}$ is the computing cost
resulting from using this particular segment list, $\CC_{1}=1$.}
\EndFor
\State $x\gets$\Call{Max}{$\gudcost$}
\State $\seglist_{j}\gets$\Call{AddSegment}{$x$}
\State \Call{RemoveUsedSFTs}{$\seglist_{j}$}
\EndWhile
\end{algorithmic}
\label{alg:greedy-compact-data-selection}
\end{algorithm}

\newpage
\bibliography{pprealopt}

\end{document}